\documentclass[a4paper,twocolumn,aps,superscriptaddress,floatfix]{revtex4}

\usepackage{amssymb,amsmath,stmaryrd}
\usepackage{times}
\usepackage{float}
\usepackage{amsmath}
\usepackage{pst-node}
\usepackage{graphicx}
\usepackage{latexsym}
\usepackage{amsfonts}
\usepackage{amssymb}
\usepackage[nouppercase]{scrpage2}
\usepackage{makeidx}
\usepackage{bbm}
\usepackage{cancel}
\usepackage{amsthm}
\usepackage{textgreek}

\newcommand{\be}{\begin{equation}}
\newcommand{\ee}{\end{equation}}
\newcommand{\bea}{\begin{eqnarray}}
\newcommand{\eea}{\end{eqnarray}}

\def\a{\alpha}
\def\b{\beta}

\def\d{\delta}
\def\D{\Delta}
\def\e{\epsilon}

\def\th{\theta}

\def\l{\lambda}

\def\m{\mu}

\def\c{\xi}

\def\p{\pi}

\def\r{\rho}
\def\s{\sigma}
\def\S{\Sigma}
\def\t{\tau}

\def\vf{\varphi}
\def\F{\Phi}

\def\w{\omega}
\def\W{\Omega}
\def\q{\psi}
\def\Q{\Psi}

\def\ble{{\mathbf e}}

\def\blD{{\mathbf D}}

\def\callN{\mbox{$\mathcal{N}$}}

\def\callT{\mbox{$\mathcal{T}$}}

\def\callZ{\mbox{$\mathcal{Z}$}}

\def\iif{\infty}
\def\bra{\langle}
\def\ket{\rangle}

\def\Tr{{\rm Tr}}
\def\Re{{\rm Re}}
\def\Im{{\rm Im}}

\def\1op{\hat{\mathbbm{1}}}
\def\nn{\nonumber}

\begin{document}

\title{Pump driven normal-to-excitonic insulator transition: 
Josephson oscillations and signatures of BEC-BCS crossover  
 in time-resolved ARPES}

\author{E. Perfetto}
\affiliation{Dipartimento di Fisica, Universit\`{a} di Roma Tor Vergata,
Via della Ricerca Scientifica 1, 00133 Rome, Italy}
\affiliation{Istituto di Struttura della Materia of the National Research 
Council, Via Salaria Km 29.3, I-00016 Montelibretti, Italy}
\author{D. Sangalli}
\affiliation{Istituto di Struttura della Materia of the National Research 
Council, Via Salaria Km 29.3, I-00016 Montelibretti, Italy}
\author{A. Marini}
\affiliation{Istituto di Struttura della Materia of the National Research 
Council, Via Salaria Km 29.3, I-00016 Montelibretti, Italy}
\author{G. Stefanucci}
\affiliation{Dipartimento di Fisica, Universit\`{a} di Roma Tor Vergata,
Via della Ricerca Scientifica 1, 00133 Rome, Italy}
\affiliation{INFN, Laboratori Nazionali di Frascati, Via E. Fermi 40, 00044 Frascati, 
Italy}


\begin{abstract}
We consider a ground-state wide-gap band insulator turning into 
a nonequilibrium excitonic insulator (NEQ-EI) upon visiting properly 
selected 
and physically relevant highly excited states.  
The NEQ-EI phase, characterized by 
 self-sustained oscillations of the complex order 
parameter, neatly follows  from a Nonequilibrium Green's Function 
treatment on the Konstantinov-Perel' contour.  We present the first 
{\em ab initio} band structure of LiF, a ground-state bulk insulator, 
in different NEQ-EI states and show that these 
states can be generated by 
currently available pump pulses.
 We highlight two general features of time-resolved 
ARPES spectra: (1) during the pump-driving  the excitonic 
spectral structure undergoes a convex-to-concave shape transition and  
{\em concomitantly} the state of the system goes through 
a BEC-BCS crossover;
(2) attosecond pulses 
shone after the pump-driving  at different times $t_{\rm delay}$ 
generate a photocurrent which 
{\em oscillates} in $t_{\rm delay}$ with a 
pump-tunable 
frequency -- we show that this phenomenon is similar to the AC response of an 
exotic Josephson junction.

\end{abstract}

\maketitle

\section{Introduction}
\label{introsec}

At low enough temperature small gap insulators, semimetals as well as metals with overlapping bands 
may turn into Excitonic Insulators (EI) due to the Coulomb electron-hole 
attraction~\cite{Blatt1962,KeldyshKopaev1965,Kozlov-Maksimov_JETP1965,JeromeRiceKohn1967,HalperinRice1968,Keldysh-Kozlov_JETP1968,Comte-Nozieres_1982}. 
Theoretical and experimental works on different materials, e.g., 
quantum Hall 
bilayers~\cite{Spielman_PhysRevLett.87.036803,Eisenstein2004},  
graphene 
bilayers~\cite{Zhang_PhysRevB.77.233405,Min_PhysRevB.78.121401,Lozovik_JETP2008,Kharitonov_PhysRevB.78.241401}, carbon nanotubes~\cite{VarsanoNC2017,Hellgren_PhysRevB.98.201103} and 
chalcogenide-based 
structures~\cite{Cercellier-PhysRevLett.99.146403,Monney_NJP2012,Zenker_PhysRevB.88.075138,Zenker_PhysRevB.90.195118,Kogar2017}, 
have addressed signatures and properties of this {\em equilibrium} EI 
phase, including 
dynamical responses to external laser 
pulses~\cite{GolezPRB2016,MurakamiPRL2017,MorPRL2017,Werdehauseneaap8652,Werdehausen_2018,Tanabe_PhysRevB.98.235127} 
and adiabatic switching of electronic correlations~\cite{Tuovinen_pssb2019}.
One of the most interesting signatures of the EI phase is the flattening of the 
valence band, recently revealed in Ta$_{2}$NiSe$_{5}$ using 
Angle-Resolved Photoemission Spectroscopy (ARPES)~\cite{Wakisaka_PhysRevLett.103.026402,Seki_PhysRevB.90.155116}.

The last decade has seen a renewed interest in 
systems which are semimetals~\cite{TriolaPRB2017,PertsovaPRB2018} or 
insulators~\cite{SzymaPRL2006,Hanai2016,HanaiPRB2017,Hanai2018,Becker_PhysRevB.99.035304,Yamaguchi_NJP2012,Yamaguchi_PhysRevLett.111.026404}
in the ground state but exhibit an EI phase in some (possibly pump-induced) 
{\em nonequilibrium} (NEQ) excited state. 
The optical properties of a NEQ-EI have been calculated by several authors in 
the past~\cite{Schmitt-Rink_PhysRevB.37.941,HAUG_1985,Glutsch_PhysRevB.45.5857,Chu_PhysRevB.54.5020}. 
However, it was not until 1993 that \"Ostreichand and Sch\"onhammer pointed out a crucial 
difference between a ground state EI and a NEQ-EI~\cite{Ostreich_1993}: in a NEQ-EI the 
macroscopic polarization {\em after the pump has been switched off} exhibits persistent, 
self-sustained oscillations 
with a finite, time-independent amplitude and a 
frequency which depends on the absorbed energy.
A similar dynamical excited state of matter 
has  been independently shown to emerge in mean-field approximations using an ansatz for 
the time-dependent order parameter~\cite{SzymaPRL2006}.
Here we provide an
alternative derivation based on nonequilibrium Green's 
functions (NEGF) and show the equivalence 
between the pump-induced state~\cite{Ostreich_1993} and the dynamical excited state~\cite{SzymaPRL2006}.

The stability of the NEQ-EI phase plays of course a 
crucial role in applications to, e.g., optoelectronics~\cite{Xiao2017}. 
Hanai {\em et al.} have pointed out the existence of regimes 
for which small fluctuations of the order parameter destroy the  
phase (dynamical instability)~\cite{HanaiPRB2017} and  
also studied the effects of an exciton-boson 
coupling~\cite{Hanai2016}. Hannewald {\em et al.} studied the NEQ-EI 
life-time due to dephasing~\cite{Hannewald-Bechstedt_2000}.
A beyond mean-field analysis including light-matter 
interaction and damping has confirmed the existence of the NEQ-EI 
phase~\cite{Becker_PhysRevB.99.035304}. Real-time first-principles studies  
are foreseeable in the near future for quantitative, 
material-dependent predictions. 

In this work we apply the NEGF formalism to systems which  exhibit a 
NEQ-EI phase but are normal Band Insulators 
(BI) in the ground state, and highlight the NEQ-EI fingerprints 
visible in ARPES 
spectra. We first
show that pump-pulses with a properly chosen subgap frequency
generate the same dynamical excited state~\cite{Ostreich_1993} as an excited self-consistent 
calculation. The proposed self-consistent NEGF 
theory is then implemented  in the {\em ab-initio}  Yambo 
code~\cite{MARINI20091392} and the 
spectral function of 
a LiF bulk insulator in the NEQ-EI phase is presented and discussed.
Our main findings are: (1) for sufficiently long and {\em non-resonant} 
pump-pulses 
the system undergoes a 
BEC-BCS crossover~\cite{Eagles_PhysRev.186.456,Strinati-2018} 
which causes 
a convex-to-concave shape transition in the excitonic  
structure of time-resolved ARPES spectra; 
(2) in the NEQ-EI phase the photocurrent generated by 
attosecond pulses shone at different times $t_{\rm delay}$ 
oscillates in $t_{\rm delay}$ with a 
pump-tunable 
frequency: we relate this phenomenon to the AC response of an 
exotic Josephson junction.

The paper is organized as follows. In Section~\ref{formalismsec} we 
illustrate the self-consistent NEGF approach to excited-state systems 
and derive the equations to calculate the
Green's function. In Section~\ref{phasediagramsec} we 
discuss the NEQ-EI phase diagram of a two-band model 
Hamiltonian. We determine the 
critical conduction density at the BEC-BCS crossover and
the boundary between the BI and NEQ-EI phases. We further illustrate 
how
the NEQ spectral function changes in these different regimes.
 The simplifications 
of the two-band model are relaxed in Section~\ref{LiFsec} where we 
present the spectral function of a LiF bulk insulator in the NEQ-EI phase. 
In Section~\ref{pumpdrivensec} we perform real-time simulations and 
obtain the optimal pump pulse which brings the system closest to the 
self-consistent NEGF results. The time-resolved ARPES spectrum is 
calculated in Section~\ref{trARPESsec} for different probe durations, 
highlighting the signatures of the BEC-BCS crossover and the 
Josephson-like oscillations. A summary of the main results and 
concluding remarks are contained in Section~\ref{conclusionssec}.

\section{Model Hamiltonian and excited-state Green's function}
\label{formalismsec}

To describe the main qualitative aspects of the NEQ-EI phase we 
initially discard several complications of realistic materials.  
We consider a one-dimensional spinless  insulator with 
one valence band and one conduction band separated by a direct gap of magnitude 
$\e_{g}$,
and assume that the intra-band repulsion 
between the electrons is negligible. A similar model has been 
proposed in Ref.~\onlinecite{YLU.2012} to discuss excitons within 
TDDFT and more recently in 
Refs.~\cite{GolezPRB2016,MorPRL2017,MurakamiPRL2017} to 
drive out of equilibrium systems which are EI in the ground state.
 In Section~\ref{LiFsec} we apply the NEGF theory to LiF 
and relax all the simplications of the model; multiple bands and  
valleys, intra-band and inter-band repulsion, band anistropies and 
degeneracies as well as spin-exchange effects will be all taken into 
account.  

The explicit form of the model Hamiltonian reads
\bea
\hat{H}&=&\sum_{k}(\e_{vk}\hat{v}^{\dag}_{k}\hat{v}_{k}
+\e_{ck}\hat{c}^{\dag}_{k}\hat{c}_{k})-
U_{0}\sum_{k}\hat{c}^{\dag}_{k}\hat{c}_{k}
\nn\\
&+&
\frac{1}{\callN}\sum_{k_{1}k_{2}q}U_{q}\,\hat{v}^{\dag}_{k_{1}+q}\hat{c}^{\dag}_{k_{2}-q}
\hat{c}_{k_{2}}\hat{v}_{k_{1}},
\label{minmodham}
\eea
where
$\hat{v}_{k}$ ($\hat{c}_{k}$) annihilates an electron of momentum $k$ in the 
valence (conduction) band, $U_{q}$ is the  Coulomb 
interaction and $\callN$ is the number of discretized $k$-points. 
The last term in the first row describes the interaction 
of (conduction) electrons with the positive background. 
The Hamiltonian in Eq.~(\ref{minmodham}) commutes with the number 
operator of valence and conduction electrons; hence each 
many-body eigenstate is characterized by a well defined number of 
electrons $N_{v}$ and $N_{c}$ in these bands. 
A nonvanishing value of the average $\bra 
\hat{c}^{\dag}_{k}\hat{v}_{k}\ket$ is possible only provided that several 
many-body eigenstates with different $N_{c}$ and $N_{v}$, constrained by 
$(N_{c}+N_{v})/\callN=1$, become degenerate in the thermodynamic limit. This can
happen for the ground-state multiplet (in this case the 
exciton energy $\e_{\rm x}\lesssim 0$)
and/or for excited-state multiplets.
In this work we are especially interested in the latter scenario.

We consider $\e_{g}/U_{0}$ large enough for the exciton energy to 
satisfy  $0\ll 
\e_{\rm x}\lesssim \e_{g}$. Then, the ground state is   
nondegenerate and it consists of a fully occupied valence band and a 
completely empty conduction band (consequently the ground-state 
average of the interaction  with the positive background vanishes). 
To investigate the existence of a NEQ-EI phase we use the 
NEGF formalism. In NEGF the 
fundamental quantity is the Keldysh-Green's function 
which for our model Hamiltonian reads~\cite{svl-book}
\be
G^{\a\b}_{k}(z,z')=\frac{1}{i}\frac{\Tr\left[
\callT\left\{e^{-i\int d\bar{z}\hat{H}(\bar{z})}
\hat{\q}_{\a k}(z)\hat{\q}_{\b 
k}^{\dag}(z')\right\}\right]}
{\Tr\left[
\callT\left\{e^{-i\int d\bar{z}\hat{H}(\bar{z})}
\right\}\right]}
\label{KeldyshG}
\ee
where $\hat{\q}_{\a k}=\hat{v}_{k},\hat{c}_{k}$ for $\a=v,c$. In 
Eq.~(\ref{KeldyshG}) the arguments $z,z'$ as well as the integral over 
$\bar{z}$ run on the Konstantinov-Perel' contour~\cite{kp.1961} consisting of a forward and 
backward branch $(0,\iif)\cup(\iif,0)$ joined to a vertical imaginary 
track $(0,-i\mbox{\textbeta})$ with $\mbox{\textbeta}$ the inverse temperature; $\callT$ is 
the contour ordering operator. In the absence of external fields 
$\hat{H}(z)=\hat{H}$ for $z$ on the forward or backward branches. 
If the system is initially in thermal equilibrium at chemical potential 
$\m$ then the Hamiltonian on the imaginary track is
$\hat{H}(z)\equiv\hat{H}^{\rm M}=\hat{H}-\m\hat{N}$ with $\hat{N}=\hat{N}_{v}+\hat{N}_{c}$ 
and $\hat{N}_{\a}\equiv\sum_{k}\hat{\q}^{\dag}_{\a k}\hat{\q}_{\a k}$ the 
number operator for electrons in band $\a$. This choice of 
$\hat{H}^{\rm M}$ does indeed correspond to averaging with the 
thermal equilibrium density matrix 
$\hat{\r}=e^{-\scriptsize{\mbox{\textbeta}}(\hat{H}-\m\hat{N})}/\callZ$, $\callZ$ being the 
partition function.
The resulting 
$G^{\a\b}_{k}(z,z')$ in Eq.~(\ref{KeldyshG}) is the  
Matsubara Green's function for both $z$ and $z'$ on the vertical 
track and the equilibrium real-time Green's functions 
for $z$ and $z'$ on the  forward or backward branches~\cite{svl-book}.

To calculate the Green's function in some excited state we must
average with an excited density matrix $\hat{\r}=
e^{-\scriptsize{\mbox{\textbeta}} \hat{H}^{\rm M}}/\callZ$, where $\hat{H}^{\rm M}$ is a 
self-adjoint operator with the property $[\hat{H}^{\rm 
M},\hat{H}]=0$. Here we consider 
\be
\hat{H}^{\rm M}=\hat{H}-\m_{v}\hat{N}_{v}-\mu_{c}\hat{N}_{c},
\label{exHM}
\ee
with $\mu_{v}\neq\m_{c}$. 
This $\hat{H}^{\rm M}$ commutes with $\hat{H}$  since 
$[\hat{N}_{\a},\hat{H}]=0$. 
For $U_{q}=0$ and zero temperature averaging with $\hat{\r}$ is 
equivalent to averaging over a state 
with the valence (conduction)
band populated up to the chemical potential $\m_{v}$ ($\m_{c}$). Hereafter we use 
$\hat{H}^{\rm M}$ in Eq.~(\ref{exHM}) for  the 
Hamiltonian on the imaginary track. 

The exact equation of motion for the Green's 
function in Eq.~(\ref{KeldyshG}) reads (in a $2\times 2$ matrix form)
\be
\left[i\frac{d}{dz}-h_{k}(z)\right]\!G_{k}(z,z')=\d(z,z')+\!\int \!
d\bar{z}\,
\S_{k}(z,\bar{z})G_{k}(\bar{z},z')
\label{eom}
\ee
where $\S_{k}$ is the sum of the Hartree-Fock (HF) and correlation
self-energies and $h_{k}(z)$ is given by
\bea
h_{k}(z)&=&\left(
\begin{array}{cc}
    \e_{vk} & 0 \\ 0 & \e_{ck}-U_{0}
\end{array}
\right)-\th_{|}(z)\left(\begin{array}{cc}
    \m_{v} & 0 \\ 0 & \m_{c}
\end{array}
\right)
\nn\\
&\equiv&h_{k}-\th_{|}(z)m
\label{hk(z)}
\eea
with $\th_{|}(z)=0$ if $z$ is on the forward or backward branches 
and $\th_{|}(z)=1$ if $z$ is on the imaginary track.
In the Hartree-Fock (HF) approximation we have
$\S_{k}^{\a\b}(z,z')=\d(z,z')V^{\a\b}_{k}(z)$,
where the HF potential reads
\bea
V^{\a\a}_{k}(z)&=&-\frac{i}{\callN}\sum_{q}U_{0}G^{\bar{\a}\bar{\a}}_{q}(z,z^{+}),
\label{HFpot1}\\
V^{\a\bar{\a}}_{k}(z)&=&\frac{i}{\callN}\sum_{q}U_{k-q}G^{\a\bar{\a}}_{q}(z,z^{+}),
\label{HFpot4}
\eea
with $\bar{\a}=c,v$ for $\a=v,c$, and $z^{+}$ the contour-time 
infinitesimally later than $z$.
As we have discarded the intraband interaction
the diagonal elements of $V$ are only due to the Hartree diagram 
whereas the off-diagonal ones are only due to the exchange (Fock) 
diagram. This simplification is relaxed in Section~\ref{LiFsec}.

\subsection{Matsubara Green's function}
\label{Matsubarasection}

Choosing $z=-i\t$ and $z'=-i\t'$ on the imaginary track we get the Matsubara Green's 
function $G^{\rm M}_{k}(\t,\t')\equiv G_{k}(-i\t,-i\t')$. Since the 
HF self-energy is local in time
the structure of $G^{\rm M}$ is the same as 
in the noninteracting case, the only difference being that the 
one-body noninteracting matrix
\be
h_{k}-m\equiv h_{k}^{\rm M},
\label{1phM}
\ee
see Eq.~(\ref{hk(z)}), is replaced 
by the one-body HF  matrix $h_{k}^{\rm M}+V_{k}$, with $V_{k}\equiv 
V_{k}(-i\t)$  the HF potential on the imaginary track~\cite{svl-book}. Therefore
\bea
G^{\rm M}_{k}(\t,\t')\!\!&=&\!\!-i\left[
\th(\t,\t')\bar{f}(h_{k}^{\rm M}+V_{k})-
\th(\t',\t)f(h_{k}^{\rm M}+V_{k})\right]
\nn\\
\!\!&\times&\!\! e^{-(\t-\t')(h_{k}^{\rm M}+V_{k})}
\label{Gmatz}
\eea
where $f(\w)=1/(e^{\scriptsize{\mbox{\textbeta}}\w}+1)$ is the Fermi 
function and $\bar{f}(\w)=1-f(\w)$. 
As $V_{k}$ depends on $G_{k}^{\rm M}$, see 
Eqs.~(\ref{HFpot1},\ref{HFpot4}), the problem has to be solved self-consistently.
We point out that for the Matsubara Green's function to be correctly 
antiperiodic in $\t$ and $\t'$ with period \textbeta, the population $n_{\a k}=-iG^{\a\a,\rm 
M}_{k}(\t,\t^{+})$ of band $\a=v,c$ is not an input of the self-consistency 
cycle.

Let $e_{k}^{\l}$ and $\vf_{k}^{\l}=\left(\begin{array}{c}
\vf_{vk}^{\l} \\ \vf_{ck}^{\l}
\end{array}\right)$, $\l=\pm$, be the two 
eigenvalues and eigenvectors  of 
$h_{k}^{\rm M}+V_{k}$. Without any loss of generality we choose 
$\vf_{\a k}^{\l}$ real.
Then, from Eq.~(\ref{Gmatz}),
\be
G^{\a\b,\rm M}_{k}(\t,\t^{+})=i\sum_{\l}
f(e^{\l}_{k})\vf^{\l}_{\a k}\vf^{\l}_{\b k},
\ee
and therefore the matrix elements of the HF potential in 
Eqs.~(\ref{HFpot1}-\ref{HFpot4}) can be rewritten as
\bea
V^{\a\a}_{k}&=&\frac{1}{\callN}\sum_{q}U_{0}\sum_{\l}
f(e^{\l}_{q})|\vf^{\l}_{\bar{\a}q}|^{2},
\label{HFpot1b}\\
V^{\a\bar{\a}}_{k}&=&-\frac{1}{\callN}\sum_{q}U_{k-q}\sum_{\l}
f(e^{\l}_{q})\vf^{\l}_{\a q}\vf^{\l}_{\bar{\a} q}.
\label{HFpot4b}
\eea
In terms of the quantities
\bea
\tilde{\e}_{vk}&\equiv& \e_{vk}+V_{k}^{vv}-\m_{v},
\label{tildeevk}
\\
\tilde{\e}_{ck}&\equiv& \e_{ck}-U_{0}+V_{k}^{cc}-\m_{c},
\label{tildeeck}
\\
\D_{k}&\equiv& V^{cv}_{k}=V^{vc}_{k},
\label{deltakdef}
\eea
it is straightforward to find the eigenvalues 
\be
e^{\l}_{k}=\frac{\tilde{\e}_{vk}+\tilde{\e}_{ck}+\l R}{2},
\label{eigenvalues}
\ee
with $R=\sqrt{(\tilde{\e}_{vk}-\tilde{\e}_{ck})^{2}+4\D_{k}^{2}}$, 
and eigenvectors 
\be
\vf_{k}^{\l}=
\left(\begin{array}{c}
\l\sqrt{\frac{1}{2}(1+\l\frac{\tilde{\e}_{vk}-\tilde{\e}_{ck}}{R})}
\\ {\rm sign}(\D_{k})
\sqrt{\frac{1}{2}(1-\l\frac{\tilde{\e}_{vk}-\tilde{\e}_{ck}}{R})}
\end{array}
\right),
\label{eigenvector-}
\ee
which inserted in Eqs.~(\ref{HFpot1b},\ref{HFpot4b}) provide a close 
system of equations for the HF potential. 
Henceforth we refer to $\D_{k}$ as the {\em order parameter} since 
a nonvanishing $\D_{k}$  implies that 
$\bra\hat{c}_{q}^{\dag}\hat{v}_{q}\ket\neq 0$ at least for some $q$, and hence  a 
spontaneous symmetry breaking. 

\subsection{Real-time Green's function}
\label{G<sec}

Once the Matsubara Green's function is known we can calculate the 
Green'function with both arguments on the real axis. In the HF 
approximation the lesser Green's function which solves Eq.~(\ref{eom}) reads~\cite{svl-book}
\be
G^{\a\b,<}_{k}(t,t')=i\sum_{\l}f(e_{k}^{\l})
\vf^{\l}_{\a k}(t)\vf^{\l\ast}_{\b k}(t')
\label{G<steadystate}
\ee
where the time-dependent vectors satisfy
\be
i\frac{d}{dt}\vf_{k}^{\l}(t)=\big[h_{k}+V_{k}(t)\big]\;\vf_{k}^{\l}(t),
\label{tdse}
\ee
with boundary conditions $\vf_{k}^{\l}(0)=\vf_{k}^{\l}$ [we recall 
that $h_{k}(z)=h_{k}$ for $z$ on the forward/backward branch, see 
again Eq.~(\ref{hk(z)})]. The 
time-dependent HF potential appearing in Eq.~(\ref{tdse}) can be 
calculated from Eqs.~(\ref{HFpot1}-\ref{HFpot4}) by taking into account 
that for $z=t$ on the forward/backward branch we have 
$G(z,z^{+})=G^{<}(t,t)$.

Let us show that the solution of Eq.~(\ref{tdse}) is 
\be
\vf_{k}^{\l}(t)=e^{-i(e^{\l}_{k}+\m-\s_{z}\frac{\d\m}{2})t}\vf_{k}^{\l},
\label{HFtdsol}
\ee
where $\s_{z}$ is the $2\times 2$ Pauli matrix,  
\be
\m\equiv\frac{\m_{c}+\m_{v}}{2}
\label{mu}
\ee
is the center-of-mass chemical potential and
\be
\d\m\equiv \m_{c}-\m_{v}
\label{deltamu}
\ee
is the relative one. Inserting Eq.~(\ref{HFtdsol}) in $G^{<}_{k}$ to 
evaluate the HF potential we find $V_{k}(t)=e^{i\s_{z}\frac{\d\m}{2} 
t}\,V_{k}\,e^{-i\s_{z}\frac{\d\m}{2} t}$. Taking into 
account that $h_{k}$ is diagonal we get
\be
\big[h_{k}+V_{k}(t)\big]\;\vf_{k}^{\l}(t)=
e^{-i(e^{\l}_{k}+\m-\s_{z}\frac{\d\m}{2})t}\big[h_{k}+V_{k}\big]\vf_{k}^{\l}.
\label{rhsintermediate}
\ee
Next we observe that $h_{k}=h_{k}^{\rm M}+m$, see Eq.~(\ref{1phM}), 
and that the $2\times 2$ matrix 
$m$ can be written as
$m=\m \mathbbm{1}-\frac{\d\m}{2}\,\s_{z}$.
Hence
\bea
\big[h_{k}+V_{k}\big]\vf_{k}^{\l}&=&
\big[h_{k}^{\rm M}+V_{k}+\m \mathbbm{1}-\frac{\d\m}{2}\,\s_{z}\big]\vf_{k}^{\l}
\nn\\
&=&
\big[e_{k}^{\l}+\m \mathbbm{1}-\frac{\d\m}{2}\,\s_{z}\big]\vf_{k}^{\l},
\eea
where in the last row we used that $\vf_{k}^{\l}$ are eigenvectors of 
$h_{k}^{\rm M}+V_{k}$. Thus, we can rewrite 
Eq.~(\ref{rhsintermediate}) as
\be
\big[h_{k}+V_{k}(t)\big]\;\vf_{k}^{\l}(t)=\big[e_{k}^{\l}+\m 
\mathbbm{1}-\frac{\d\m}{2}\,\s_{z}\big]\vf_{k}^{\l}(t)
\ee
which coincides with the time-derivative 
$i\frac{d}{dt}\vf_{k}^{\l}(t)$.  We 
emphasize that all quantities needed to evaluate the lesser
Green's function have been previously calculated to determine the 
Matsubara Green's function. The greater Green's function can be derived similarly; the final 
result looks like Eq.~(\ref{G<steadystate}) after replacing $if(e_{k}^{\l})\to 
-i\bar{f}(e_{k}^{\l})$. 

Due to the presence of $\s_{z}$ in Eq.~(\ref{HFtdsol}) only the diagonal elements 
of $G^{<}_{k}$ depend on the time-difference. The off-diagonal ones oscillate 
monochromatically at the frequency $\d\m/2$. In particular, 
the order parameter 
$\D_{k}(t)=V^{cv}_{k}(t)=[V^{cv}_{k}(t)]^{\ast}$ oscillates 
monochromatically at the
frequency $\d\m$, i.e.,
\be
\D_{k}(t)=\D_{k}e^{-i\d\m t}
\label{Delta(t)},
\ee
in agreement with Ref.~\cite{SzymaPRL2006}. 
The time-depencence of the order parameter in the NEQ-EI phase 
resembles the behavior of the superconducting order parameter 
in a Josephson junction. Josephson oscillations
between two superconducting electrodes  separated by a thin insulator arise 
by applying a DC voltage to the equilibrium junction, thus 
effectively introducing a difference in the electrochemical 
potentials. This mechanism has recently been 
investigated by replacing the superconductors with EI 
electrodes~\cite{Rontani_PhysRevB.80.075309,Joglekar_PhysRevB.72.205313,Hsu2015,Apinyan2019}.
The NEQ-EI oscillations in Eq.~(\ref{Delta(t)}) are of a 
slightly different nature. They 
can be understood in terms of an exotic 
Josephson junction where Cooper pairs are formed by electrons in 
different electrodes. In our system the electrodes are the valence 
and conduction band and the Cooper pairs are the bound excitons. 
As we shall see in 
Section~\ref{photocurrentsec}, the oscillating behavior of $\D_{k}$  has interesting 
implications in time-resolved photocurrent spectra.

\section{Phase diagram}
\label{phasediagramsec}

\begin{figure}[tbp]
\includegraphics[width=0.47\textwidth]{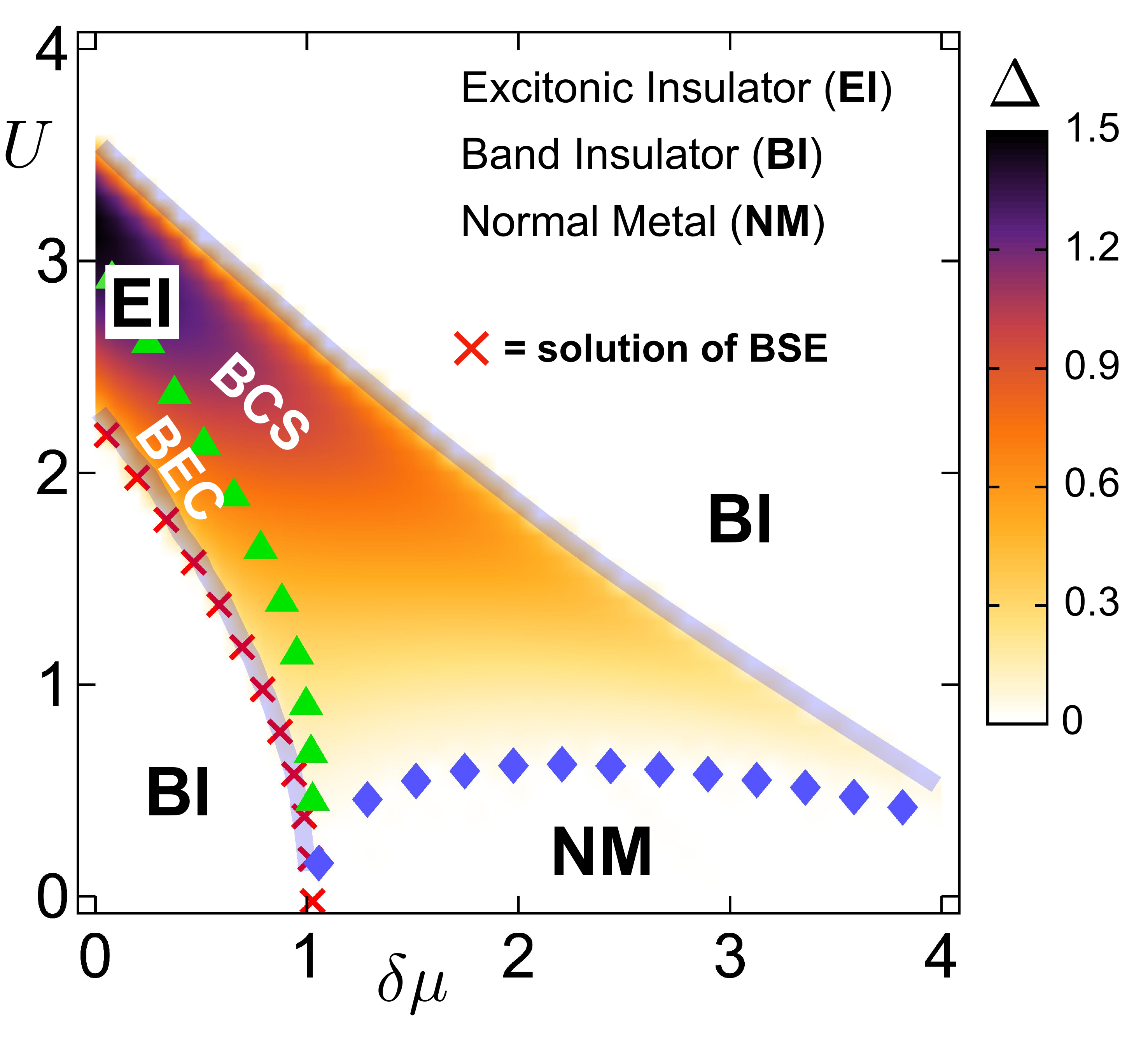}
\caption{Color plot of of $\D$ for different $\d\m$ 
and interaction strength $U$. Red crosses at the phase boundary correspond 
to the energy of the zero-momentum exciton at interaction strength $U$
(for $U=0$ this energy is the bare gap).
The value of $\D$ has been selected 
on the basis of the principle of minimum energy.
The BEC-BCS crossover (green triangle) is determined by the condition 
$\c=r_{s}$ where $\c$ is the width of the vanishing-momentum 
excitonic wavefunction in real space and $r_{s}$ is the average 
distance between electrons in the conduction band.
The EI-NM crossover (blue diamonds) is determined by the condition 
$\D<10^{-2}$. All energies are 
in units of the bare band gap $\e_{g}$.}
\label{phasediagram}
\end{figure}

To illustrate the possible solutions of the self-consistent 
calculation we consider the band structure
\be
\e_{\a k}=(-)^{\a}[2T(1-\cos(k))+\e_{g}/2]
\label{bands}
\ee
where $(-)^{v}\equiv -1$ 
and $(-)^{c}\equiv 1$, $T>0$, and a 
short-range Coulomb interaction $U_{q}=U$ independent of $q$. 
The wavevectors $k,q,\ldots$ vary in the first Brillouin zone $(-\p,\p)$.
Then 
the HF potential $V^{\a\b}_{k}$ is independent of $k$, see 
Eqs.~(\ref{HFpot1b},\ref{HFpot4b}), and hence 
$\D_{k}=\D$ is independent of $k$ too. We consider the system at zero 
temperature and express all energies in units 
of the bare gap $\e_{g}$. We fix the bandwidths $W=4T=2$, the 
center-of-mass chemical potential $\m=0$
and vary $U\geq 0$ and $\d\m\geq 0$.
For $\d\m=0$ we recover the ground-state Green's function.
Figure~\ref{phasediagram} shows the color-plot of the 
self-consistent $\D$. In the white regions 
we have the trivial (and unique) solution $\D=0$. These 
are the BI regions characterized by $N_{c}= 0$ or $N_{v}=0$, and hence by a gap 
$\e_{g}$ between occupied an unoccupied  
single-particle states. In the colored 
region we have two solutions, the trivial one ($\D=0$) and the nontrivial 
one ($\D\neq 0$). For both solutions we have estimated the total 
energy according to~\cite{svl-book}
\be
E=-\frac{i}{2}\sum_{k}\int\frac{d\w}{2\p}\Tr\left[(\w+h_{k})G^{<}_{k}(\w)\right]
\ee
and found that the energy of the nontrivial solution is always 
the lowest. In Section~\ref{specfuncsec} we calculate the spectral 
function and show that the unoccupied and occupied single-particle states 
are separated by an energy gap $\D$. Hence the system is still an insulator. However, since 
$N_{c}$ and $N_{v}$ are both nonvanishing this insulating phase can only be due to the Coulomb 
attraction between a conduction electron and a valence hole. We then 
say that the system is in the EI phase. The EI 
phase exists in the ground state ($\d\m=0$) for large enough $U$ as well as in excited 
states ($\d\m\neq 0$) even for rather small $U$'s. The region below the blue diamonds is 
characterized by a nontrivial solution with small $\D$ (in this 
region $\D\leq 10^{-2}$);
here the system behaves essentially like a normal metal (NM).

\subsection{Spectral function}
\label{specfuncsec}

\begin{figure*}[tbp]
\includegraphics[width=0.9\textwidth]{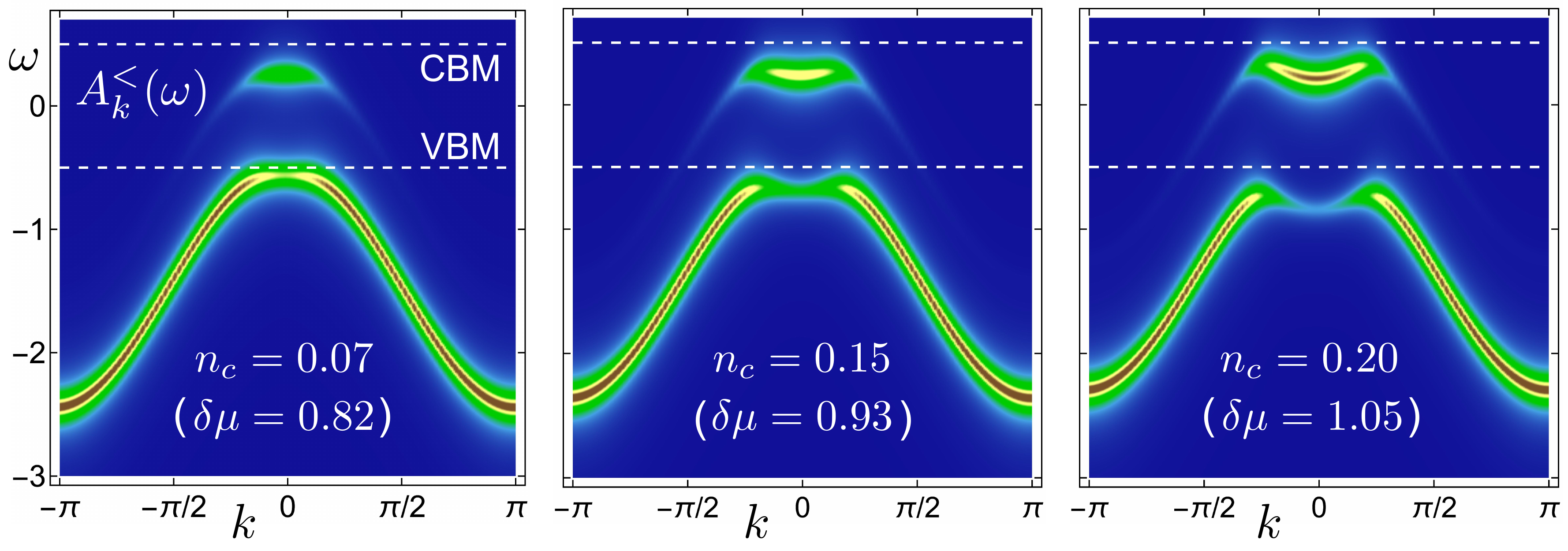}
\caption{Occupied part of the spectral function for different values 
of $\d\m$ and $U=1$. Dashed lines indicate the conduction band 
minimum (CBM) and valence band maximum (VBM). All energies are in units of the bare gap $\e_{g}$.}
\label{SCspectral}
\end{figure*} 

From the NEGF soluton of Section~\ref{G<sec} we can extract the occupied part of the spectral 
function according to
\be
A^{<}_{k}(\w)=-i\Tr[G^{<}_{k}(\w)]
\ee
where $G^{\a\a,<}_{k}(\w)$ is the Fourier transform of 
$G^{\a\a,<}_{k}(t,t')$  with respect to the time-difference. In 
Fig.~\ref{SCspectral} we show results for $U=1$ and three different 
$\d\m=0.82,\,0.93,\,1.05$ corresponding to a
density of 
electrons in the conduction band  
$n_{c}=-i\sum_{k}G^{cc,<}_{k}(t,t)/\callN=0.07,0.15,\,020$.
 Notice that for $U=1$ the 
nontrivial solution $\D\neq 0$ exists only for values of 
$\d\m\gtrsim 0.76$,
see again Fig.~\ref{phasediagram}.
Interestingly, for our parameters the energy of the zero-momentum exciton is 
$\e_{\rm x}\simeq 0.76$. The fact that $\e_{\rm x}$ coincides with the value of 
$\d\m$ at the BI-EI boundary is not a 
coincidence~\cite{Yamaguchi_NJP2012}. For completeness we provide a 
proof in Appendix~\ref{NI-EIboundarysec}.
For $\d\m=0.82$ ({\em low} densities $n_{c}$) an excitonic structure appears at roughly $\e_{g}-\d\m$
below the conduction band minimum (CBM). This spectral function agrees with the 
results of Ref.~\cite{RustagiKemper2018} where 
a system with one {\em single} exciton, i.e., $n_{c}=1/\callN$, was 
considered.

By increasing further $\d\m$ (and hence the density in the 
conduction band) the excitonic structure changes its convexity, see 
middle and right panel in Fig.~\ref{SCspectral}. This 
phenomenon is distinct from the one reported in 
Ref.~\cite{RustagiKemper2018} where the change of convexity is 
obtained by  
averaging spectral functions each with one single exciton of 
different center-of-mass momentum. In our case the convex-to-concave 
shape transition 
develops with increasing the density of excitons, 
see also Section~\ref{BEC-BCSsec}. 

The unoccupied part of the spectral function can be calculated 
similarly: $A^{>}_{k}(\w)=-i\Tr[G^{>}_{k}(\w)]$. Since 
$\m=0$ the system is particle-hole symmetric and therefore 
\be
A^{>}_{k}(\w)=A^{<}_{k}(-\w).
\ee
As anticipated there is a gap of order $\D$ between the unoccupied 
and occupied bands.

\subsection{BEC-BCS crossover}
\label{BEC-BCSsec}
The EI phase is similar to the BCS superconducting phase when  
the width $\c$ of the excitons is larger than the average electron distance (or 
Wigner-Seitz radius) $r_{s}$. In the opposite regime, i.e., $\c\ll 
r_{s}$, the excitons behave like point-like bosons in a 
Bose-Einstein condensate (BEC). To determine the 
BEC-BCS 
crossover~\cite{Comte-Nozieres_1982,Nozieres1985,Yamaguchi_NJP2012,Strinati-2018} we have calculated the width $\c$ of the zero-momentum 
excitonic wavefunction and compare it with the average distance $r_{s}$ of electrons in the 
conduction band. The excitonic wavefunction in real space is  
given by (modulo a normalization constant)
$Y(x)=\sum_{k}e^{ikx}Y_{k}$,
with $Y_{k}$ the excitonic solution of Eq.~(\ref{bcs}), whereas (for 
one-dimensional systems) $r_{s}\simeq \callN/N_{c}$.
The green triangles in the phase diagram correspond to the values of 
$U$ and $\d\m$ for which $\c=r_{s}$, where $\c$ has been estimated 
from $Y(\c)/Y(0)=0.1$.

\begin{figure}[tbp]
\includegraphics[width=0.4\textwidth]{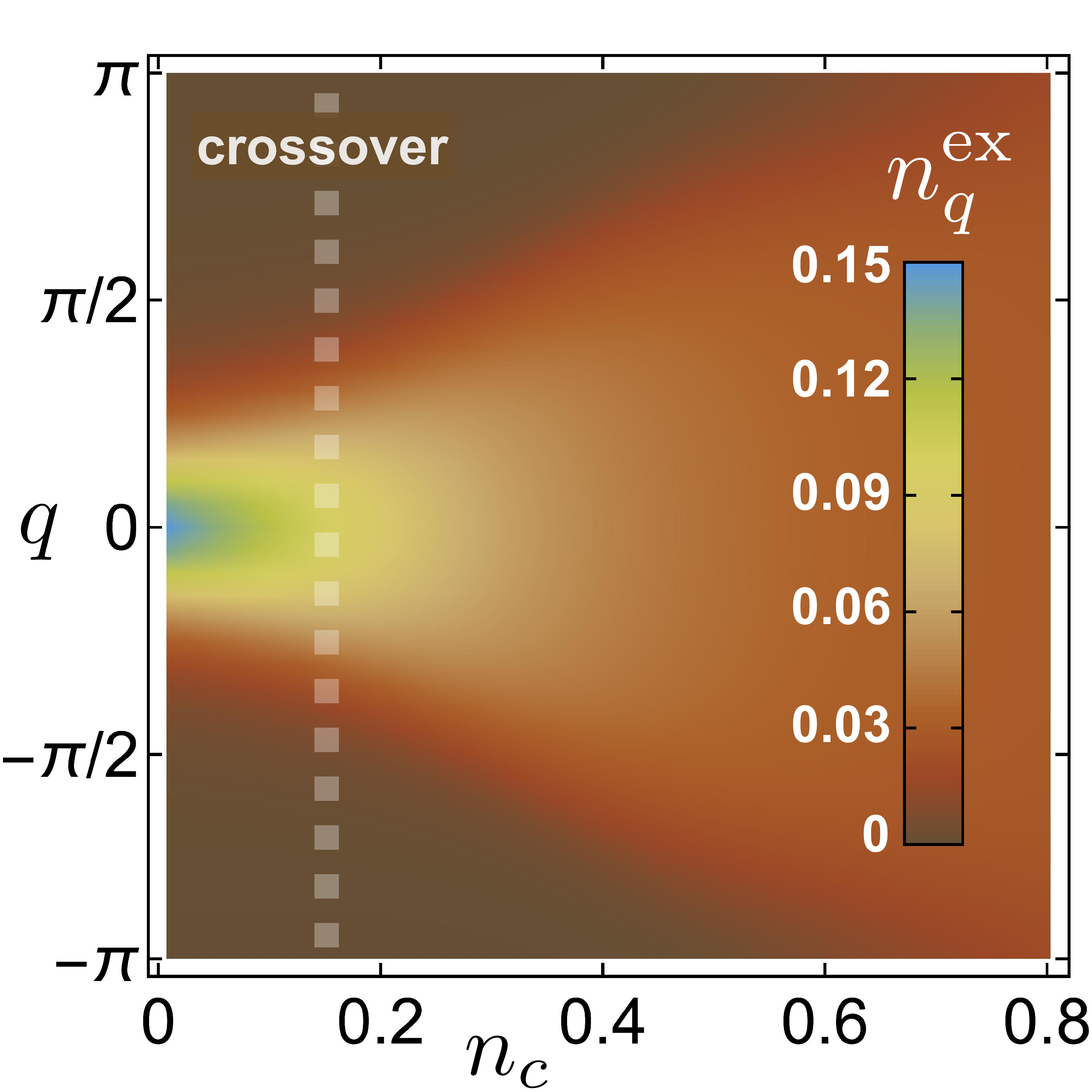}
\caption{Color plot of the excitonic population in momentum space 
for different densities $n_{c}=N_{c}/\callN=1/r_{s}$ for $U=1$.}
\label{excitondensity}
\end{figure} 

An alternative way to find the values of $\d\m$ at the BEC-BCS crossover 
consists in determining the onset of the broadening of the excitonic 
population in momentum space -- in 
the BEC regime the excitonic population is peaked around  
zero momentum.
Let us write the many-body state of the system in the NEQ-EI phase. 
Since the minus branch $e_{k}^{-}$ is entirely below 
the plus branch $e_{k}^{+}$ we have
\be
|\Q_{\rm x}\ket=\prod_p(\vf_{vp}^{-} 
\hat{v}^\dag_p+\vf_{cp}^{-}\hat{c}^\dag_p)|0\ket,
\ee
where $|0\ket$ is the state with no electrons.
We define the creation operator for an exciton of momentum $q$ 
according to
\be
\hat{b}^\dag_q=\sum_k Y_k^{(q)}\hat{c}^\dag_{k+q}\hat{v}_k
\ee
where the excitonic amplitude $Y_{k}^{(q)}$ can be calculated by 
solving the analogous of Eq.~(\ref{bcs}) for finite-momentum 
excitons. It is a matter of simple algebra to show that
\be
\bra\Q_{\rm x}|\hat{c}^{\dag}_{k'+q'}\hat{v}_{k'}
\hat{v}_k^{\dag}\hat{c}_{k+q}|\Q_{\rm 
x}\ket=\d_{qq'}\d_{kk'}(\vf_{ck}^{-}\vf_{ck+q}^{-})^{2}.
\ee
Therefore, the excitonic density in the NEQ-EI phase is given by
\be
n_{q}^{\rm ex}\equiv \bra\Q_{\rm x}|\hat{b}^{\dag}_{q}\hat{b}_{q}|\Q_{\rm 
x}\ket=\sum_k(Y_k^{(q)}\vf_{ck}^{-}\vf_{ck+q}^{-})^2.
\ee
In Fig.~\ref{excitondensity} we show $n_{q}^{\rm ex}$ 
for the interaction strength $U=1$ and different 
densities in the conduction band. The broadening starts 
for $n_{c}=N_{c}/\callN=1/r_{s}\simeq 0.15$. This density 
is obtained when $\d\m\simeq 0.93$,
in agreement with the abscissa of the green 
triangle in Fig.~\ref{phasediagram} for the same value of $U$.

Remarkably, the convex-to-concave shape transition  in the 
excitonic structure of the spectral function  
roughly occurs at the BEC-BCS crossover. In Fig.~\ref{SCspectral} 
the Coulomb strength is $U=1$ and the transition occurs for 
$\d\m\simeq 0.93$, in 
agreement with the previously estimated value. We have 
verified (not shown) that this property remains true for all values 
of $U<U_{c}\simeq 2.3$ (for larger $U$'s the ground state is an EI). 
We infer that time-resolved ARPES spectra may  
provide a tool to detect the BEC-BCS crossover in NEQ-EI's, see also 
Section~\ref{BECBCSARPESsec}.

\section{LiF bulk insulator}
\label{LiFsec}

We now apply the NEGF theory for excited states, see 
Section~\ref{formalismsec}, to a LiF bulk insulator. 
In the ground state this material has an experimental gap 
of about 14.5 eV and the optical 
spectrum exhibits a $1s$ exciton with a binding energy of
$1.9$~eV~\cite{Roessler:67}.

To describe the system we have chosen the Kohn-Sham (KS) basis of a 
self-consistent calculation performed with
the QuantumEspresso package~\cite{QuantumEspresso} at the level of
the local density approximation (LDA). 
The ground-state has been converged using a $6\times 6\times 6$, $\Gamma$ centered
grid with a cutoff of $80$ Ry. Then the wave-functions have been 
computed with a non self-consistent calculation on a $8\times 8\times 8$ grid. 
Our equilibrium bands well reproduce those of Ref.~\cite{Wang2003} 
with a KS gap of $8.45$ eV, which is quite far from the experimental value. 
To improve the description we have 
constructed the Hartree plus Screened EXchange (HSEX) 
potential using the Yambo code~\cite{MARINI20091392}. 
Multiple bands and  
valleys, intra-band and inter-band repulsion, band anistropies and 
degeneracies as well as spin-exchange effects are all built in.  
A first approximated screened 
interaction has been evaluated in RPA with KS energies and wave-functions for the 
polarization. The HSEX spectrum is found to be well approximated 
by the original KS spectrum if the KS valence (conduction) energies are 
stretched by a factor 1.65 (1.4). The stretched KS energies 
with a scissor of 6~eV (necessary to reach the experimental gap)
have been used to improve the 
polarization and hence the screened interaction. 
We verified that a 
further iteration of the screened interaction did not change 
substantially the stretching factors. 
The HSEX with improved interaction widens the gap from 8.5~eV to 
12.8~eV; we have therefore applied a scissor of 1.7~eV to match the 
experimental value of the gap.
The improved screened interaction has also been implemented in the 
Bethe--Salpeter equation with kernel given by the functional 
derivative of the HSEX potential. The optical spectrum well reproduces the experimental results.
In particular the $1s$ excitonic peak is located at about $\e_{\rm 
x}=12.48$~eV with respect to the VBM, hence its binding 
energy is $2$~eV.

Having a satisfactory description  of LiF in equilibrium we have 
carried out the self-consistent Matsubara procedure of 
Section~\ref{formalismsec} using the HSEX potential with ground-state RPA 
screened interaction. 
For both 
equilibrium and excited-state calculations  
we have explicitely considered 
50 bands. The screened interaction and the screened contribution to the 
exchange term has been calculated with $8$~Ry ($=59$ RL vectors) 
whereas the Hartree term and the unscreened contribution 
to the exchange term has been calculated with  
$64$~Ry ($965$ RL vectors). 
In the self-consistency cycles we have fixed the total charge per 
unit cell (charge neutrality) instead of 
the center-of-mass chemical potential $\m$.
Convergence paramenters for the HSEX calculation are chosen to have 
an error $<0.1~$eV.

\begin{figure}[tbp]
\includegraphics[width=0.48\textwidth]{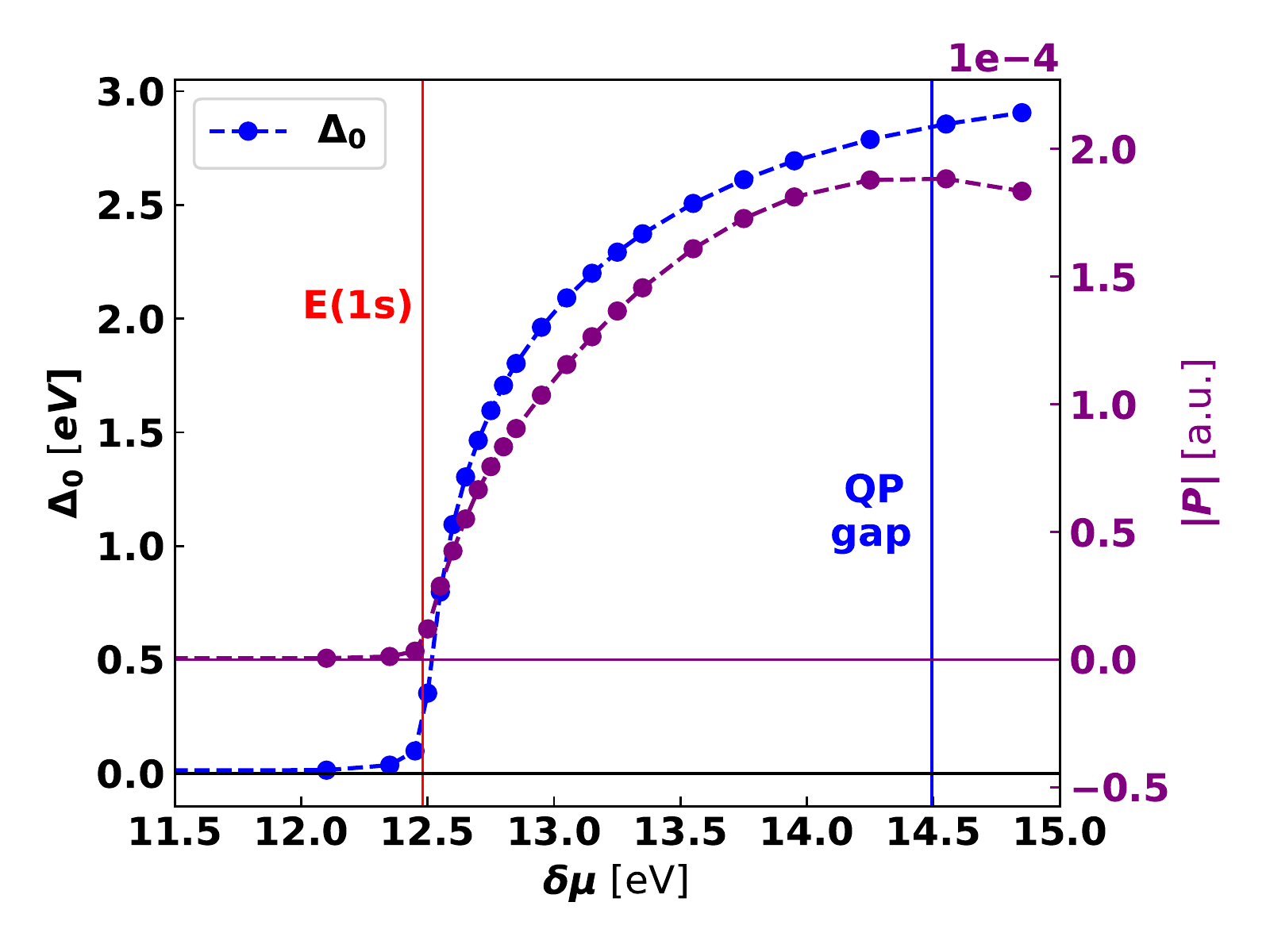}
\caption{Modulus of 
the order parameter (blue) and macroscopic 
polarization (purple) versus $\d\m$ for a bulk LiF. The zero of 
energy is the VBM. The vertical 
lines have been drawn in correspondence of the energy of the 1$s$ 
exciton (red) and quasi-particle gap (blue).}
\label{fig:phasediagram_LiF}
\end{figure} 
\begin{figure*}[tbp]
\includegraphics[width=0.322\textwidth]{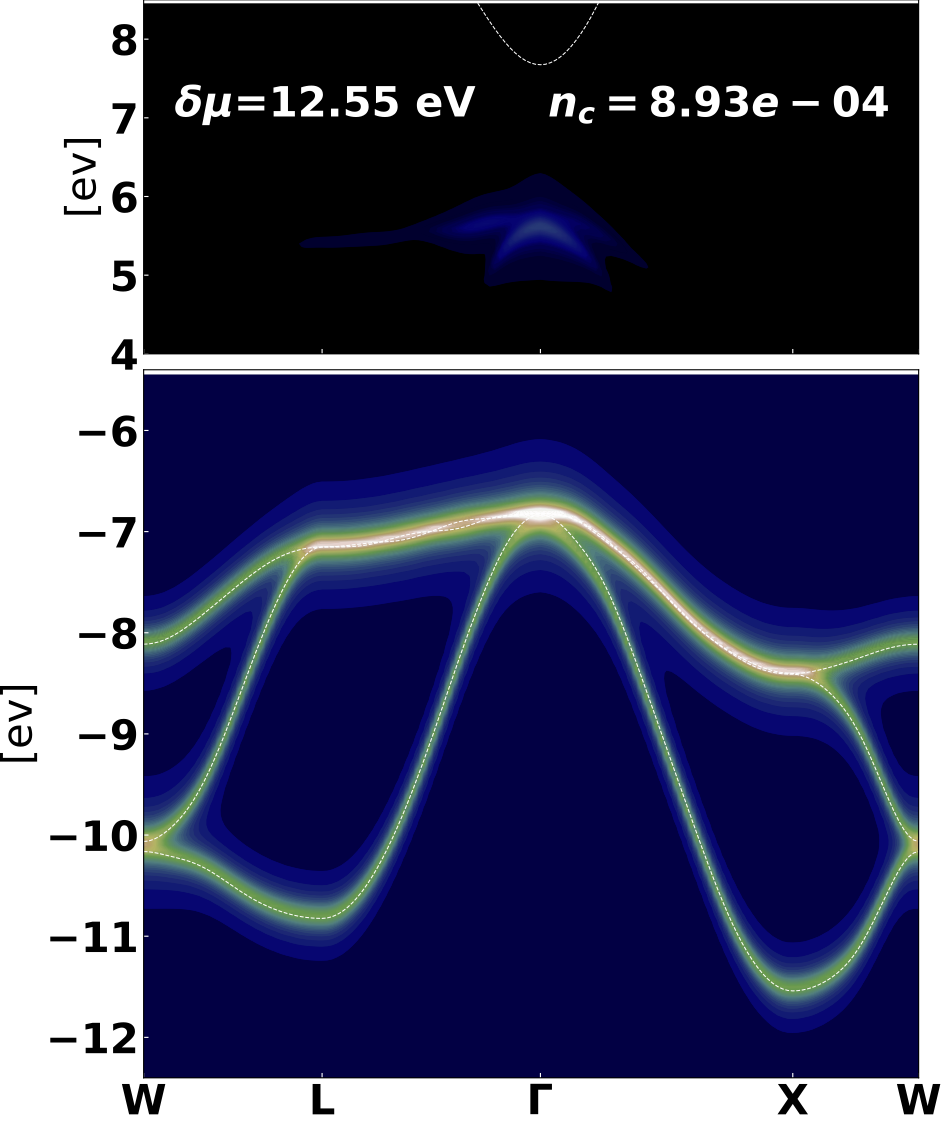}
\includegraphics[width=0.281\textwidth]{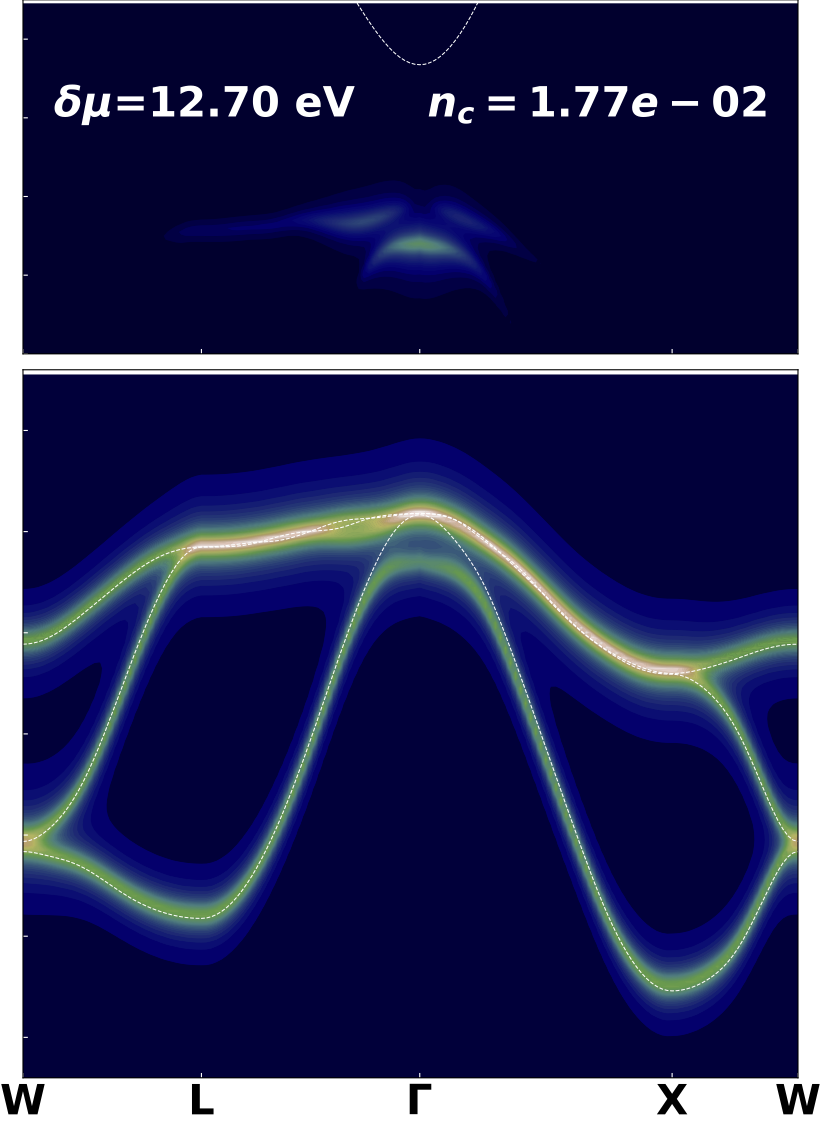}
\includegraphics[width=0.334\textwidth]{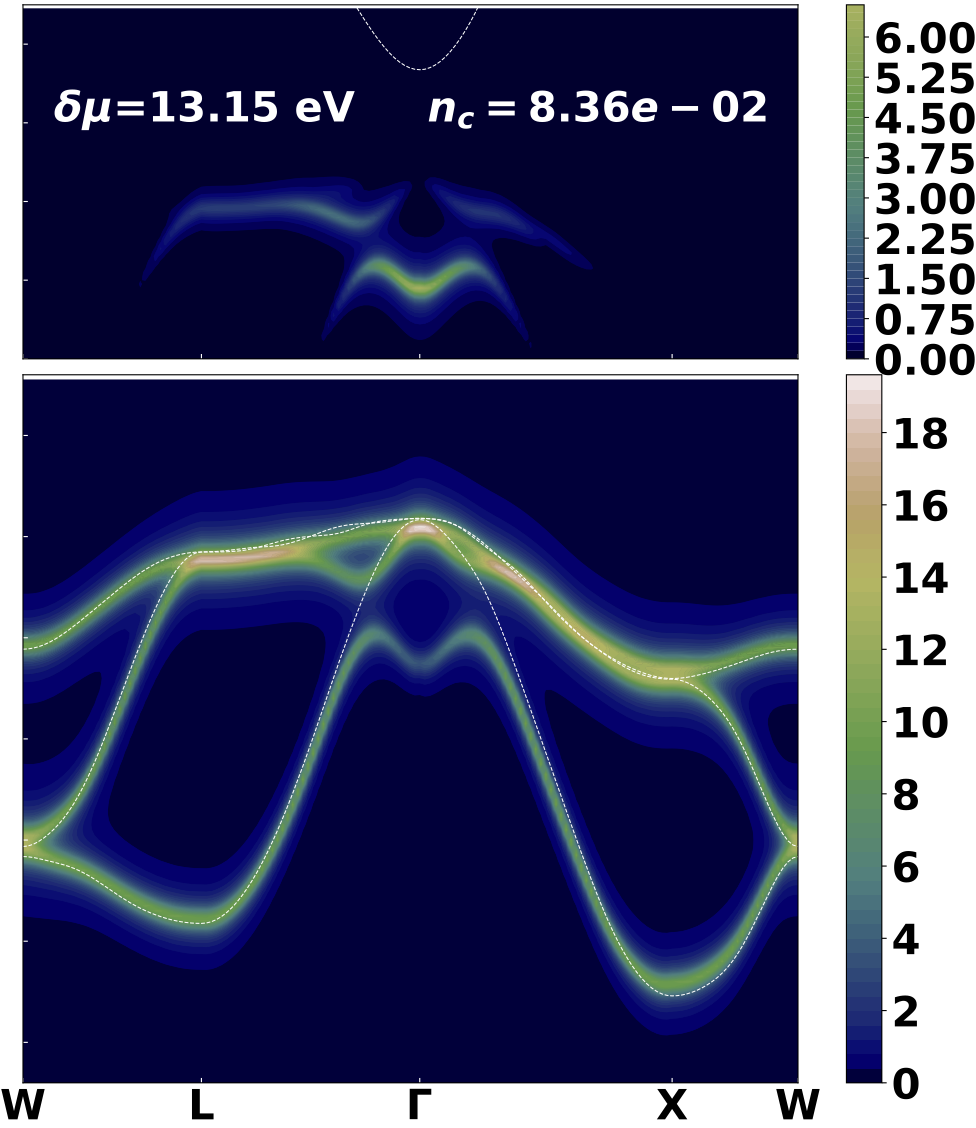}
\caption{Occupied part of the spectral function of a bulk LiF for different values of $\d\m$.
         The values of the chemical potential and of the resulting density in the
         conduction band are reported in the figure. The white dashed lines represents
         the equilibrium band structure.}
\label{SCspectralLiF}
\end{figure*} 

In Fig.~\ref{fig:phasediagram_LiF} we show the order parameter $\D_{0}$ 
and the macroscopic polarization 
$P$ versus $\d\m$. In LiF we have defined the order parameter as the 
following zero-momentum average
\be
\D_k=\sum_{v,c} h^{\rm HSEX}_{cvk},
\label{deltakdef_ai}
\ee
where $h^{\rm HSEX}$ is the HSEX Hamiltonian, and the sum extends 
over all valence ($v$) and conduction ($c$) 
bands. 
Notice that Eq.~\eqref{deltakdef_ai} reduces to Eq.~\eqref{deltakdef}
in the two-band model if 
the HSEX hamiltonian is replaced by the HF one.  
By construction the order parameter $\D_{k}$ is basis-dependent.
However, the transition to the NEQ-EI phase can also be detected by 
looking at the macroscopic polarization $P$, which is a measurable 
physical quantity and, as shown in Fig.~\ref{fig:phasediagram_LiF}, 
a good order parameter 
($\D$ and $P$ are proportional in the two-band model system).
For $\d\m<\e_{\rm x}$ both $\D$ and $P$ vanish, in 
agreement with the findings of the previous Section.

LiF 
shares other important qualitative features with the model 
Hamiltonian. In Fig.~\ref{SCspectralLiF} we 
display the occupied part of the spectral function for three different values of $n_{c}$.
Both the development of the excitonic structure and
the convex-to-concave shape transition with increasing $n_{c}$ are 
clearly visible. 
We also observe that excitons distribute among different bands giving 
rise to multiple excitonic peaks at fixed parallel momentum $k$.

The results presented in this Section
support the use of the two-band model for understanding other general 
features of the NEQ-EI phase.

\section{Pump driven BI-EI transition}
\label{pumpdrivensec}

In this Section we demonstrate that the self-consistent NEQ-EI phase 
is accessible by shining suitable laser pulses on the BI ground 
state.

We consider again the Hamiltonian of Eq.~(\ref{minmodham}) at 
zero temperature
and set $\d\m=0$ (ground-state) 
and $U<U_{c}$ (BI phase with $\D=0$).
 The system is driven out of equilibrium by a time-dependent electric field $E(t)$
coupled to the valence-conduction dipole moments $D_{k}$. The driving 
Hamiltonian reads
\be
\hat{H}_{\rm 
drive}(t)=E(t)\sum_{k}D_{k}(\hat{c}^{\dag}_{k}\hat{v}_{k}+\hat{v}^{\dag}_{k}\hat{c}_{k}).
\label{Hdrive}
\ee
This light-matter coupling has been used to study 
changes in the {\em optical} properties due to a 
strong monochromatic electric field (dynamical Stark 
effects)~\cite{Comte_PhysRevB.34.7164,Comte_PhysRevB.38.10517}
and more recently weak 
monochromatic fields as those generated in quantum optical cavities~\cite{Latini_QED-BSE}.   
Here we are interested in the {\em electronic} properties of the system 
when the electric field is a pump-pulse centered 
around frequency $\w_{P}$ with finite duration $T_{P}$:
\be
E(t)=\th(1-|1-2t/T_{P}|)
E_{P}\sin^{2}(\frac{\p t}{T_{P}})\sin(\w_{P}t).
\ee
All results in this Section refer to a system with interaction strength 
$U=1$, dispersion of Eq.~(\ref{bands}) with bandwidth $W=4T=2$, 
center-of-mass chemical potential 
$\m=0$ and, for 
simplicity, dipole moments $D_{k}=D$ independent of $k$ 
(energies are in units of the bare gap $\e_{g}$). The real-time 
simulations have been performed with the CHEERS code~\cite{PS-cheers}.
Since the driving Hamiltonian depends on $E_{P}$ and $D$ only through 
their product we define the Rabi frequency $\W_{P}\equiv E_{P}D$.

\begin{figure}[tbp]
\includegraphics[width=0.47\textwidth]{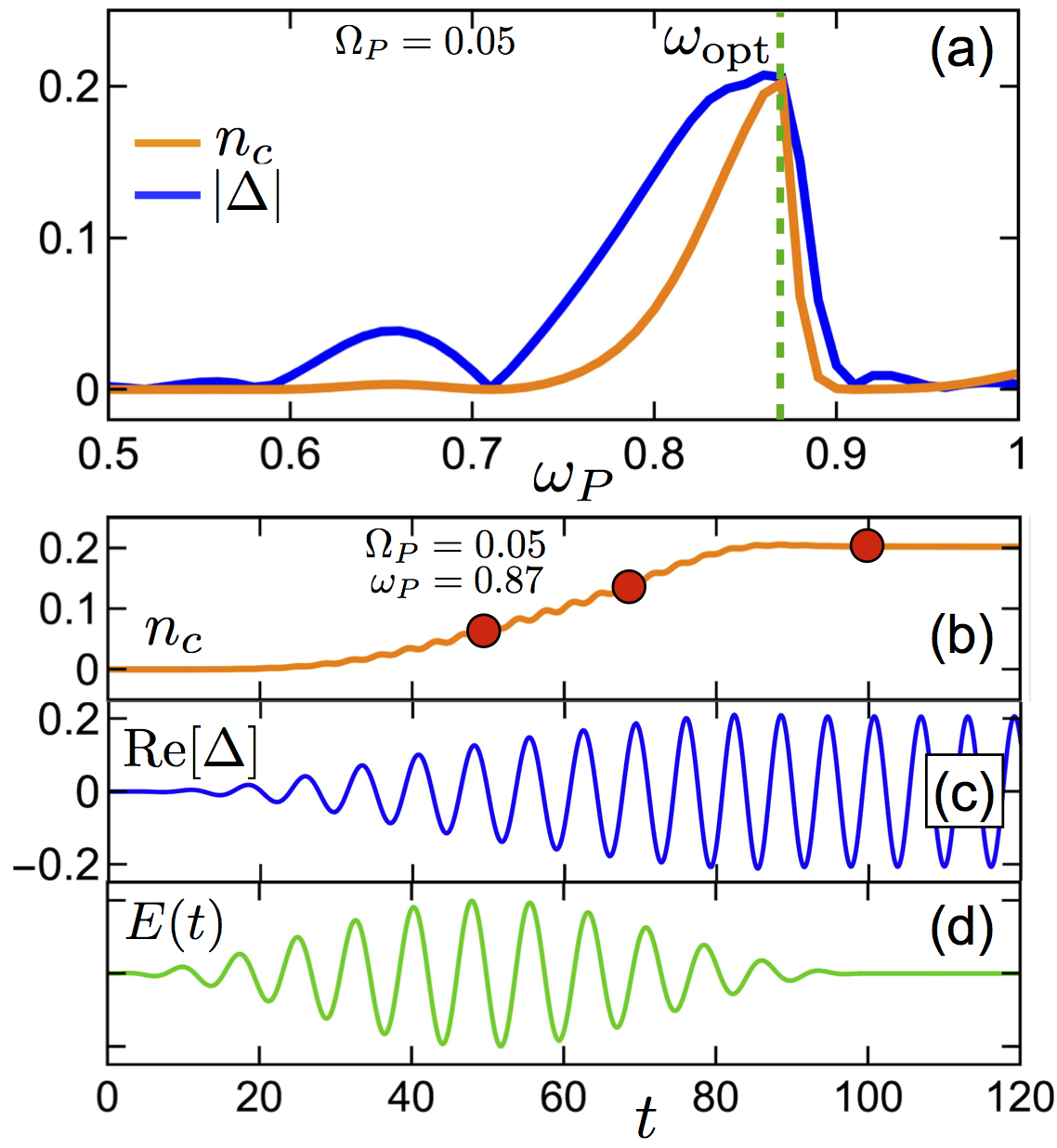}
\caption{(a) Density in the conduction band $n_{c}$ and modulus of the order 
parameter $|\D|$ after the pump versus the pump frequency 
$\w_{P}$. Time-dependent density in the conduction band (b), real 
part of the order parameter (c) and pump-pulse profile (d). 
The red dots in panel (b) are drawn at the values of 
$n_{c}=0.07,\,0.15,\,0.20$. 
Energies are in units of $\e_{g}$ and times in units of 
$1/\e_{g}$.}
\label{TD1}
\end{figure} 

We evolve the system in the time-dependent HF approximation and 
express times in units of $1/\e_{g}$.
The main result of the simulations is that for $\w_{P}$ larger than 
the exciton energy $\e_{\rm x}$, see Eq.~(\ref{bcs}), and smaller than
$\e_{g}$ {\em the modulus of the order 
parameter attains a constant and finite value after the pump}, in 
agreement with Ref.~\cite{Ostreich_1993}. 
In Fig.~\ref{TD1}(a) we show the 
steady values of the density in the conduction band $n_{c}$ and 
of the modulus 
of the order 
parameter $|\D|$ versus the pump frequency 
$\w_{P}$ (we recall that for the chosen parameters $\e_{\rm x}=0.76$) 
for a pump-pulse of duration $T_{P}=100$ and Rabi 
frequency $\W_{P}=0.05$, see Fig.~\ref{TD1}(d). Independently of the values of $T_{P}$ and $\W_{P}$ the 
trend of Fig.~\ref{TD1}(a) is general: there exists an {\em optimal frequency} 
$\w_{\rm opt}\in(\e_{\rm x},\e_{g})$ (of course depending on $T_{P}$ and $\W_{P}$) at which 
the steady values of $n_{c}$ and $|\D|$ are simultaneously 
maximized. We point out that in the noninteracting case, $U=0$, the steady value of $|\D|$ is 
zero for all $\w_{P}$, $T_{P}$ and $\W_{P}$ due to perfect dephasing.

In Fig.~\ref{TD1}(b-c) we show $n_{c}(t)$ and 
$\Re[\D(t)]$ during the action of the pump-pulse with $\w_{P}=\w_{\rm 
opt}\simeq 0.87$. 
We observe the occurrence of persistent monochromatic oscillations in $\Re[\D(t)]$
after $T_{P}=100$. We therefore conclude that 
\be
\D(t>T_{P})=e^{-i\w_{E} t}|\D|,
\ee
where $\w_{E}$ is a pump-dependent frequency.
The time-dependent behavior of the order parameter is the same as 
that of the self-consistent solution, see Eq.~(\ref{Delta(t)}). 
We have extracted the frequency 
$\w_{E}$ from the real-time solution $\Re[\D(t)]$, set $\d\m=\w_{E}$ in the 
self-consistent calculation of Section~\ref{Matsubarasection} and 
found that {\em the self-consistent values of $n_{c}$ and $\D$ 
are identical to the steady-state values of $n_{c}$ and $|\D|$}. 
This remains true for several different pump intensities (not shown).
We therefore conclude that the pump-driven state is precisely the NEQ-EI state 
analyzed in detail in Section~\ref{formalismsec}.

\begin{figure}[tbp]
\includegraphics[width=0.4\textwidth]{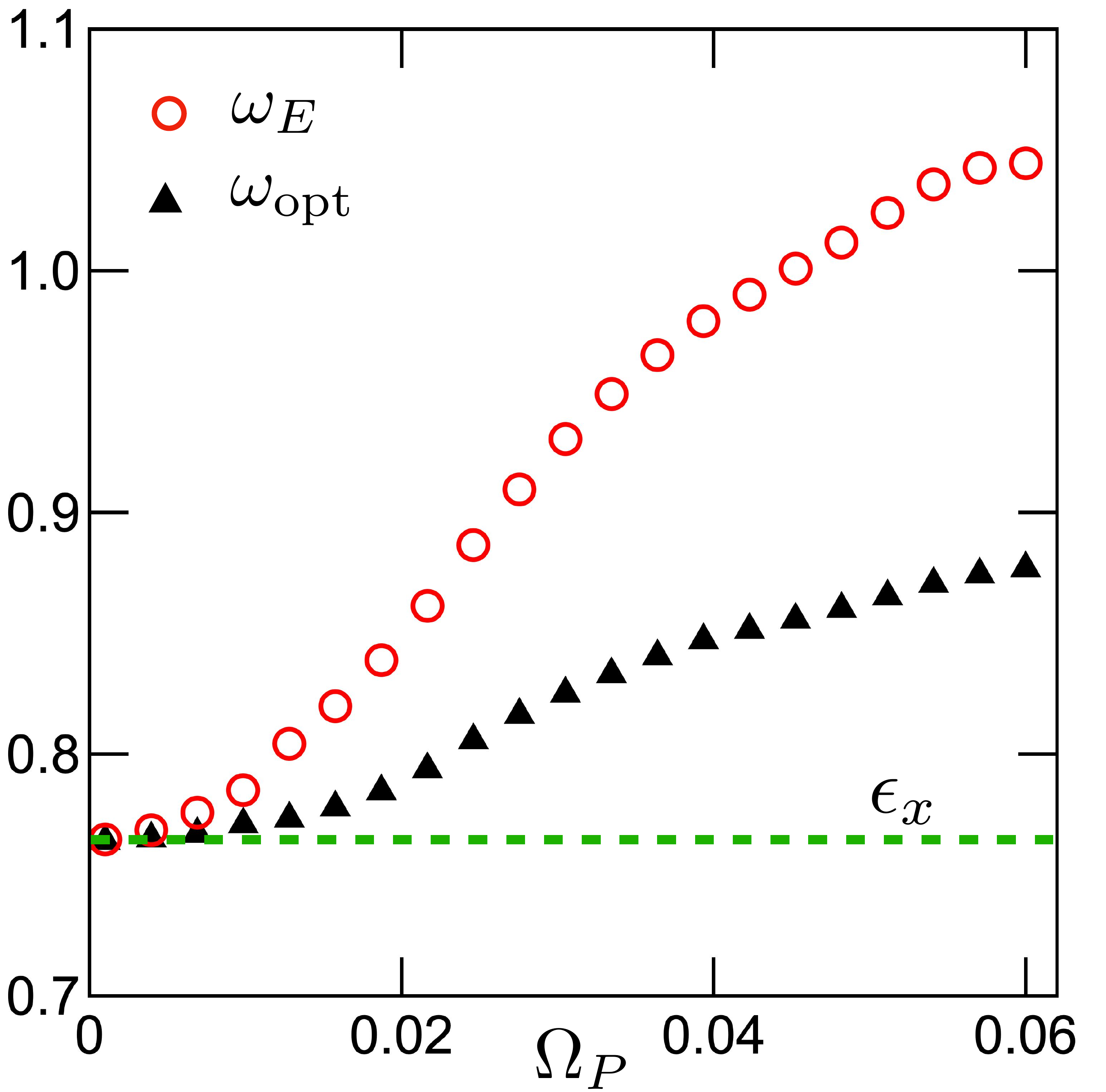}
\caption{Frequency $\w_{E}$ (empty circle) resulting from 
pumping at the optimal frequency (black triangle)  for different 
Rabi frequencies $\W_{P}$. The value of the excitonic energy $\e_{\rm 
x}=0.76$ 
is highlighted with a dashed green line. All energies are in units of $\e_{g}$.}
\label{TD2}
\end{figure} 

A second evidence in favor of our conclusion is presented in 
Fig.~\ref{TD2} where we show the values of $\w_{E}$ (empty circle) resulting from 
pumping at the optimal frequency (black triangle)  for $T_{P}=100$ 
and different 
Rabi frequencies $\W_{P}$. In the limit of vanishing $\W_{P}$ (hence 
vanishing intensities of the electric field) the 
density pumped in the conduction band approaches zero and both $\w_{E}$ 
and $\w_{\rm opt}$ approach the energy $\e_{\rm x}=0.76$ of the 
zero-momentum exciton (dashed green line). This is the same behavior 
of $\d\m$ versus $n_{c}$. Indeed, 
a vanishingly small value of $n_{c}$ in the self-consistent solution 
implies a vanishingly small value of $\D$ which we know to occur for
$\d\m=\e_{\rm x}$, see Appendix~\ref{NI-EIboundarysec}.

\section{Time-resolved ARPES}
\label{trARPESsec}

The time-resolved ARPES signal is proportional to the 
number of electrons $N_{k}(\w)$  with 
energy $\w$ and parallel momentum $k$
ejected by a probing pulse $\ble(t)$. 
For arbitrary probe pulses we have~\cite{PSMS.2016,Freericks_PhysRevLett.102.136401}
\be
N_{k}(\w)=2\sum_{\a\b}\int dt d\bar{t}\;\Re\left[
\S^{\a\b,\rm R}_{k,\w}(t,\bar{t})G^{\b\a,<}_{k}(\bar{t},t)\right],
\label{Nkw}
\ee
where the ionization self-energy reads
\be
\S^{\a\b,\rm R}_{k,\w}(t,\bar{t})=-i\th(t-\bar{t})
[\blD_{k,\w}^{\a}\cdot\ble(t)]e^{-i\w(t-\bar{t})}
[\blD_{k,\w}^{\b}\cdot\ble(\bar{t})]
\ee
and $\blD_{k,\w}^{\a}$ is the dipole matrix element between a band state 
$\a k$ and a continuum LEED state of energy  $\w$ and parallel momentum 
$k$. 

\subsection{BEC-BCS crossover with femtosecond probe-pulses}
\label{BECBCSARPESsec}

\begin{figure*}[tbp]
\includegraphics[width=0.9\textwidth]{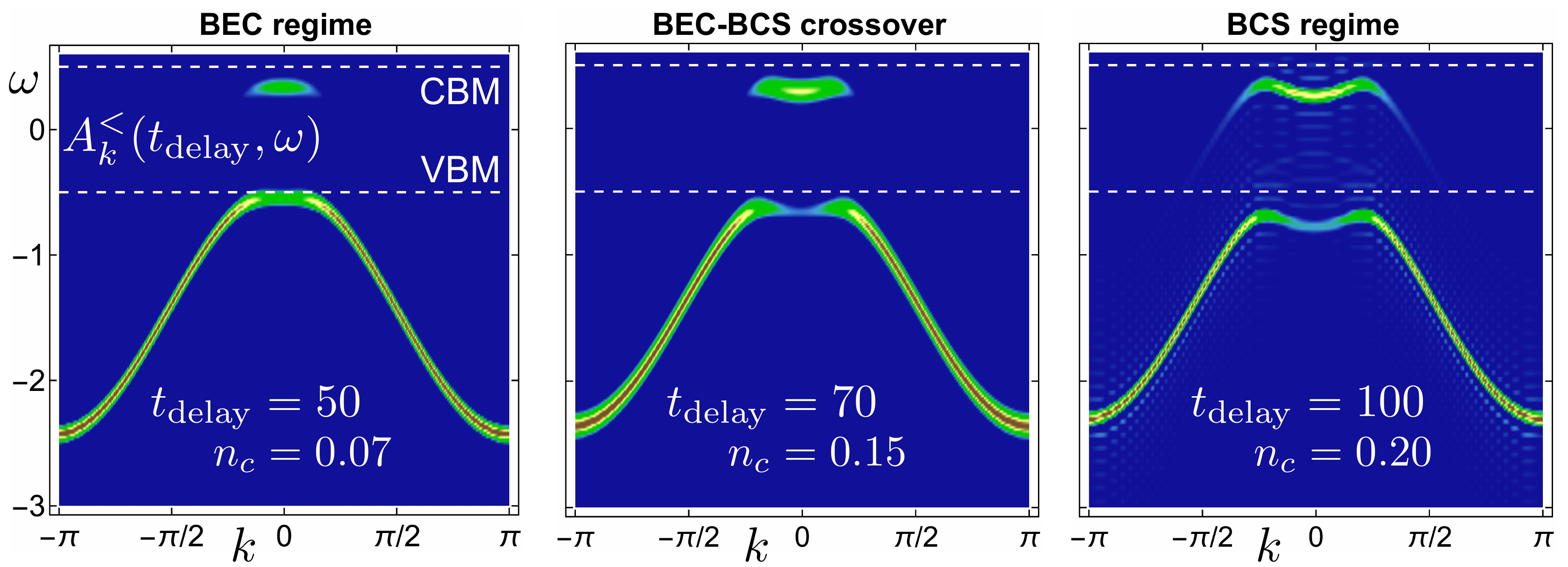}
\caption{Occupied part of the transient spectral function for 
different delays corresponding to a density in the conduction band 
$n_{c}=0.07,\,0.15,\,0.20$. 
Energies are in units of $\e_{g}$ and times in units of $1/\e_{g}$.}
\label{TDspectral}
\end{figure*} 

From Eq.~(\ref{Nkw}) we see that $N_{k}(\w)$ is a complicated 
two-times convolution of $G^{<}(t,t')$. Let us introduce the relative 
time $\t\equiv t-t'$ and the center-of-mass time 
$t_{\rm CM}\equiv (t+t')/2$. Since the lesser Green's function yields the 
probability amplitude for a hole to propagate freely  
from $t'$ to $t$ the $G^{<}(t,t')$ varies in $\t$ no slower 
than the inverse of the energy of the highest occupied state. For a system 
in equilibrium this energy is the ionization potential; in our 
nonequilibrium system this energy can be estimated as the CBM, i.e., 
$\e_{c0}$. 
The time-scale of the variation of $G^{<}$ in the center-of-mass time 
$t_{\rm CM}$ 
can be inferred from the rate of change of the occupations, see 
Fig.~\ref{TD1}(b), and can be estimated as a fraction of the 
pump duration $T_{P}$. For band gaps in the eV range our pump pulse has a duration 
$T_{P}\simeq 
10^{2}$~fs. Therefore, a probe pulse of duration $T_{p}\simeq 
T_{P}/10\simeq 10^{1}$~fs and  
central frequency $\w_{p}\gg 2\p/T_{p}$ is enough to resolve  
all removal energies provided that $T_{p}\gg 2\p/\e_{c0}$. If such a  
probe impinges on the system at time $t_{\rm delay}$ then 
$N_{k}(\w)\propto A^{<}_{k}(t_{\rm delay},\w-\w_{p})$ where the 
transient spectral function is defined according to
\be
A^{<}(t_{\rm delay},\w)=-i\int d\t \,e^{i\w\t}\,\Tr\left[
G^{<}(t,t')\right]
\label{transientSF}
\ee
and $t_{\rm delay}=t_{\rm CM}$. Of course, for $t_{\rm delay}\gg T_{P}$ the 
transient spectral function becomes independent of $t_{\rm delay}$.

To obtain $A^{<}(t_{\rm delay},\w)$ we observe that
the time-dependent HF equations give access 
to the ``time diagonal'' $G_{k}^{<}(t,t)$ from which we 
can calculate the retarded HF Green's function 
\be
G^{\rm R}_{k}(t,t')=-i\th(t,t')\callT\left\{
e^{-i\int_{t'}^{t}d\bar{t}[h_{k}+D_{k}E(\bar{t})\s_{x}+V_{k}(\bar{t})]}\right\}
\ee
where $\s_{x}$ is the Pauli matrix, see Eq.~(\ref{Hdrive}).
From $G^{\rm R}_{k}(t,t')=[G^{\rm A}_{k}(t',t)]^{\ast}$ we can extract
the ``time off-diagonal'' $G_{k}^{<}(t,t')$ as~\cite{svl-book}
\be
G_{k}^{<}(t,t')=-iG^{\rm R}_{k}(t,t')G_{k}^{<}(t',t')
+iG_{k}^{<}(t,t)G^{\rm A}(t,t'),
\ee
and calculate the transient spectral function from 
Eq.~(\ref{transientSF}).

We consider again the pump-pulse in Fig.~\ref{TD1}(d).
The green dots are drawn at times $t=50,\,70,\,100$
for which $n_{c}=0.07\,,0.15,\,0.20$ respectively; these values of the 
conduction density are the same as in the three panels of 
Fig.~\ref{SCspectral}. In Fig.~\ref{TDspectral} we show the color plot of 
$A^{<}(t_{\rm delay},\w)$ at $t_{\rm delay}=50,70,100$. For  
$t_{\rm delay}\lesssim T_{P}$ the pump is still active and the transient spectral function 
is affected by nonadiabatic effects~\cite{Myohanen_2010}. 
However, these are very small and the agreement between Fig.~\ref{SCspectral} 
and Fig.~\ref{TDspectral} is unexpectedly good. 
Incidentally, we observe that this agreement supports 
the adiabatic approximation employed in Ref.~\cite{PSMS.2016} 
to calculate time-resolved ARPES spectra. 
For $t_{\rm delay}\gg T_{P}$ the
transient spectral function  becomes independent of $t_{\rm 
delay}$ (as it should be) and almost
indistinguishable from the right panel of Fig.~\ref{SCspectral} (not 
shown). This 
corroborates further the equivalence between the pump-driven state and the 
NEQ-EI state of  Section~\ref{formalismsec}.

\begin{figure}[tbp]
\includegraphics[width=0.47\textwidth]{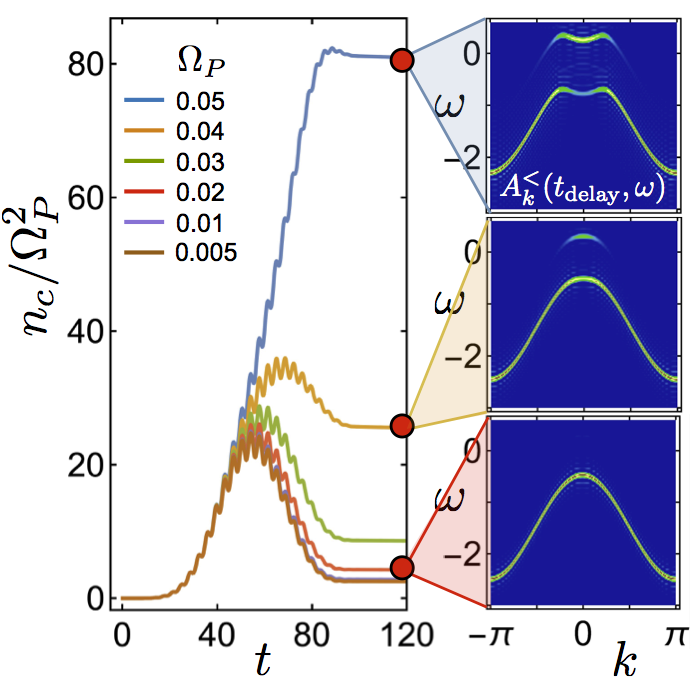}
\caption{Time-dependent density in the conduction band for pump 
pulses of duration $T_{P}=100$, central frequency 
$\w_{P}=0.87$ and different Rabi frequencies $\W_{P}$. The 
right panels show the steady-state spectral function calculated at 
$t_{\rm delay}=110$ for 
$\W_{P}=0.02,\,0.04,\,0.05$. Energies are in units of $\e_{g}$ and 
times in units of $1/e_{g}$.}
\label{linear-analysis}
\end{figure}

From Fig.~\ref{TDspectral} we also see that the convex-to-concave 
shape transition of the excitonic structure (signaling the BEC-BCS 
crossover) can be revealed by time-resolved ARPES spectra. 
It is worth emphasizing that linear response theory, i.e., 
$n_{c}\propto \W_{P}^{2}$, is totally inadequate to describe the 
pump-driven evolution of the system toward 
the BCS domain.  In 
Fig.~\ref{linear-analysis} we show the time-dependent evolution of 
$n_{c}(t)$ for $\w_{P}=0.87$, $T_{P}=100$ and different values of the 
Rabi frequency $\W_{P}$. For $\W_{P}=0.02,\,0.04,\,0.05$ we display 
the corresponding steady-state ($t_{\rm delay}\gg T_{P}$) spectral 
functions. The concavity changes for  $\W_{P}=0.05$ and for this 
Rabi frequency the steady density of conduction electrons is clearly not proportional to 
$\W_{P}^{2}$.  

\subsection{Josephson oscillations with attosecond  probe-pulses}
\label{photocurrentsec}

The ARPES signal changes dramatically for ultrashort probe pulses 
since the probe has no time to wash out the oscillatory contribution 
of the off-diagonal $G^{<}$. 
To highlight the main qualitative difference we consider a
$\d$-like pulse of the form
\be
\ble(t)=\ble_{0}\,\d(t-t_{\rm delay}).
\ee
Then, the number of electrons in Eq.~(\ref{Nkw}) becomes
\be
N_{k}(\w)=-\sum_{\a\b}\Im\left[\W^{\a}_{k,\w}\W^{\b}_{k,\w}
G^{\b\a,<}_{k}(t_{\rm delay},t_{\rm delay})\right]
\ee
where we have defined the Rabi frequencies 
\be
\W^{\a}_{k,\w}=\blD_{k,\w}^{\a}\cdot \ble_{0}.
\ee
At the steady state 
$G^{\a\a,<}_{k}(t,t)$ is independent of $t$ but 
$G^{cv,<}_{k}(t,t)=-[G^{vc,<}_{k}(t,t)]^{\ast}\sim e^{-i\d\m t}$, see 
Eq.~(\ref{G<steadystate}). 
Consequently $N_{k}(\w)$ consists of a DC part $N^{\rm (DC)}_{k}(\w)$ 
and an AC part of amplitude $N^{\rm (AC)}_{k}(\w)$ and frequency 
$\d\m$:
\be
N_{k}(\w)=N^{\rm (DC)}_{k}(\w)+N^{\rm (AC)}_{k}(\w)\sin[(\d\m) t_{\rm 
delay}+\vf].
\label{ACtrARPES}
\ee
We therefore predict {\em oscillations} in $N_{k}(\w)$ versus 
$t_{\rm delay}$ {\em even for} $t_{\rm delay}\gg T_{P}$.
These oscillations should be observed in time-resolved ARPES 
provided that the duration of the probe pulse is much smaller 
than the period $2\p/\d\m$. For values of $\d\m$ in the eV 
range (like in LiF) probe durations of the order of  a hundred of 
attoseconds are 
sufficient.

The AC response in Eq.~(\ref{ACtrARPES})
can again be explained 
using the analogy with the exotic Josephson junction, see Section~\ref{G<sec}.
The application of a pump pulse coupling electrons in different 
electrodes (bands in our case) is  equivalent to the application of a 
DC bias between the electrodes. If the DC bias is kept on for a time $T_{P}$ then 
a macroscopic number of electrons is transferred from one electrode 
(valence band) to the other (conduction band) thus generating a 
difference in the electrochemical potentials of the electrodes. Once the 
DC bias (pump in our case) is switched off this difference does not damp to zero  since electrons 
cannot hop back to the original electrode (band). By all means this difference in 
electrochemical potentials is equivalent to the difference $\d\m$ 
discussed in Section~\ref{formalismsec}.

\section{Summary and outlook}
\label{conclusionssec}

We presented a derivation of the NEQ-EI phase 
using the NEGF formalism on the 
Konstantinov-Perel' contour. The NEQ-EI phase can be physically 
induced by pump-pulses with properly chosen subgap frequency, and the nature 
(BEC or BCS) of the excitonic condensate can be tuned by the 
pump duration. The most remarkable difference between a ground-state 
EI and a NEQ-EI is the time-dependent behavior of the order 
parameter and the total polarization. In fact, they both exhibit 
persistent, self-sustained monochromatic oscillations 
even at vanishing pump. Interestingly, these oscillations have the 
same nature of the AC oscillations in an exotic Josephson junction, 
where Cooper pairs are formed by electrons at the opposite side of the 
junction.

The NEQ-EI phase leaves clear fingerprints in time-resolved ARPES spectra. 
For probe durations long-enough to resolve the band structure the oscillatory 
contribution of the order parameter is washed out 
and an excitonic ``band''
appears inside the gap. Depending on the BEC or BCS nature of the 
excitonic condensate this band can be either convex or concave. 
Time-resolved ARPES experiments could in principle monitor the 
convex-to-concave shape transition using, e.g., pump-pulses of 
hundreds of femtoseconds and overlapping probe-pulses of few 
femtoseconds shone at different delays. Ultrafast probe pulses of, e.g., 
a few hundreds of attoseconds, do instead broaden the band structure 
and unveil the time-dependent contribution.
The resulting photocurrent 
oscillates in time even after the end of the 
pump. The period of the oscillations is the same as the period of the 
order parameter, which is no smaller than the exciton energy $\e_{\rm 
x}$.

The equivalence between the time-dependent approach and the 
self-consistent NEGF scheme for excited states allows for studying 
the NEQ-EI phase in realistic materials using different strategies. 
The inclusion of electronic correlations at, e.g., the GW level is 
more feasible in the self-consistent scheme. On the other 
hand, the effects of phonon-induced coherence-losses and exciton 
recombination on the self-sustained oscillations of the total 
polarization could be investigated using the real-time Kadanoff-Baym 
equations~\cite{svl-book} in the GKBA framework~\cite{PhysRevB.34.6933}. 
Of course accurate calculations are needed for quantitative 
predictions on specific materials. However, the highlighted qualitative 
features of the NEQ-EI phase are general and provide a useful guide 
for the 
interpretion of time-resolved ARPES spectra.

\appendix

\section{Bethe-Salpeter equation and the BI-EI phase boundary}
\label{NI-EIboundarysec}

The equation for the  BI-EI boundary marked with red crosses in 
Fig.~\ref{phasediagram} can be 
obtained analytically.
Infinitesimally close to the boundary $\D_{k}$ 
is small. Then the minus band $e^{-}_{k}$ is entirely below the 
plus band $e^{+}_{k}$ and only $\l=-$ contributes in 
the sum of Eq.~(\ref{HFpot4b}). 
Expanding Eq.~(\ref{eigenvector-}) to lowest order in $\D_{k}$ and 
taking into account that $\tilde{\e}_{vk}-\tilde{\e}_{ck}<0$
we find
\be
\vf_{vk}^{-}\simeq -1,
\quad\quad
\vf_{ck}^{-}\simeq- \frac{\D_{k}}{\tilde{\e}_{vk}-\tilde{\e}_{ck}}.
\label{wklow}
\ee
Using this expansion the diagonal elements of 
the HF potential can be approximated to first order in $\D_{k}$ 
as $V^{vv}\simeq 0$ and $V^{cc}_{k}\simeq U_{0}$, see 
Eqs.~(\ref{HFpot1b}). Consequently, from 
Eqs.~(\ref{tildeevk},\ref{tildeeck}) we get $\tilde{\e}_{vk}\simeq 
\e_{vk}-\m_{v}$ and $\tilde{\e}_{ck}\simeq \e_{ck}-\m_{c}$, which 
implies that $\tilde{\e}_{vk}-\tilde{\e}_{ck}=\e_{vk}-\e_{ck}+\d\m$. 
Using this result in Eq.~(\ref{wklow}), inserting the resulting 
expressions  into Eq.~(\ref{HFpot4b}) and recalling that, by the 
definition in Eq.~(\ref{deltakdef}), 
$V^{vc}_{k}=\D_{k}$  we get
\be
\D_{k}=-\frac{1}{\callN}\sum_{q}U_{k-q}\frac{\D_{q}}{\e_{vq}-\e_{cq}+\d\m}.
\label{NIEIboundaryeq}
\ee
The lowest value of $\d\m$ for which this equation admits nontrivial 
solutions defines the BI-EI boundary. 

Interestingly, Eq.~(\ref{NIEIboundaryeq})
coincides with the eigenvalue equation of a many-body eigenstate 
with a single exciton of vanishing momentum. 
Let $|\Q_{0}\ket=\prod_{k}\hat{v}^{\dag}_{k}|0\ket$ be
the ground state of energy $E_{0}$ in the BI phase and let us 
write the one-exciton state as
\be
|\Q\ket=\sum_{k}Y_{k}\hat{c}^{\dag}_{k}\hat{v}_{k}|\Q_{0}\ket=
\sum_{k}Y_{k}|\F_{k}\ket,
\label{exexp}
\ee
where we introduced the {\em eh} states 
$|\F_{k}\ket\equiv\hat{c}^{\dag}_{k}\hat{v}_{k}|\Q_{0}\ket$.
Since $\hat{H}$ preserves the number of electrons in each band 
$\hat{H}|\Q\ket$ is again a linear combination of the 
$|\F_{k}\ket$'s. The possible excited state energies 
$E=E_{0}+\e_{\rm x}$ are found by solving the eigenvalue problem 
$\hat{H}|\Q\ket=(E_{0}+\e_{\rm x})|\Q\ket$. Inserting the expansion in 
Eq.~(\ref{exexp}) and taking the sandwich with $\bra\F_{k}|$ we get 
the Bethe-Salpeter equation for the coefficients $Y_{k}$
\be
(\e_{ck}-\e_{vk}-\e_{\rm x})Y_{k}=\frac{1}{\callN}\sum_{q}U_{k-q}Y_{q}.
\label{bcs}
\ee
Renaming $\e_{\rm x}=\d\m$ and setting 
$Y_{q}=\D_{q}/(\e_{vq}-\e_{cq}+\d\m)$ we see by inspection that Eq.~(\ref{bcs}) is 
identical to Eq.~(\ref{NIEIboundaryeq}). Thus, for a given $U$ the 
BI-EI boundary is found at $\d\m=\e_{\rm x}$, in agreement with 
Ref.~\cite{Yamaguchi_NJP2012}.
For $U=0$ the BI-EI boundary is at $\d\m=\e_{g}$, as it should, 
whereas for $U\geq U_{c}\simeq 2.3$ 
(in this case the ground state is an EI)
the solutions of Eq.~(\ref{NIEIboundaryeq}) occur for 
$\d\m\leq 0$.
The $\d\m\leq 0$ criterion is indeed used to
establish the occurrence of an equilibrium EI 
phase~\cite{HalperinRice1968}.


\begin{thebibliography}{68}
\expandafter\ifx\csname natexlab\endcsname\relax\def\natexlab#1{#1}\fi
\expandafter\ifx\csname bibnamefont\endcsname\relax
  \def\bibnamefont#1{#1}\fi
\expandafter\ifx\csname bibfnamefont\endcsname\relax
  \def\bibfnamefont#1{#1}\fi
\expandafter\ifx\csname citenamefont\endcsname\relax
  \def\citenamefont#1{#1}\fi
\expandafter\ifx\csname url\endcsname\relax
  \def\url#1{\texttt{#1}}\fi
\expandafter\ifx\csname urlprefix\endcsname\relax\def\urlprefix{URL }\fi
\providecommand{\bibinfo}[2]{#2}
\providecommand{\eprint}[2][]{\url{#2}}

\bibitem[{\citenamefont{Blatt et~al.}(1962)\citenamefont{Blatt, B\"oer, and
  Brandt}}]{Blatt1962}
\bibinfo{author}{\bibfnamefont{J.~M.} \bibnamefont{Blatt}},
  \bibinfo{author}{\bibfnamefont{K.~W.} \bibnamefont{B\"oer}},
  \bibnamefont{and} \bibinfo{author}{\bibfnamefont{W.}~\bibnamefont{Brandt}},
  \bibinfo{journal}{Phys. Rev.} \textbf{\bibinfo{volume}{126}},
  \bibinfo{pages}{1691} (\bibinfo{year}{1962}),
  \urlprefix\url{https://link.aps.org/doi/10.1103/PhysRev.126.1691}.

\bibitem[{\citenamefont{Keldysh and Kopaev}(1965)}]{KeldyshKopaev1965}
\bibinfo{author}{\bibfnamefont{L.~V.} \bibnamefont{Keldysh}} \bibnamefont{and}
  \bibinfo{author}{\bibfnamefont{Y.~U.} \bibnamefont{Kopaev}},
  \bibinfo{journal}{Sov. Phys. Solid State} \textbf{\bibinfo{volume}{6}},
  \bibinfo{pages}{2219Ð2224} (\bibinfo{year}{1965}).

\bibitem[{\citenamefont{Kozlov and Maksimov}(1965)}]{Kozlov-Maksimov_JETP1965}
\bibinfo{author}{\bibfnamefont{A.~N.} \bibnamefont{Kozlov}} \bibnamefont{and}
  \bibinfo{author}{\bibfnamefont{L.~A.} \bibnamefont{Maksimov}},
  \bibinfo{journal}{JETP} \textbf{\bibinfo{volume}{21}}, \bibinfo{pages}{790}
  (\bibinfo{year}{1965}).

\bibitem[{\citenamefont{J\'erome et~al.}(1967)\citenamefont{J\'erome, Rice, and
  Kohn}}]{JeromeRiceKohn1967}
\bibinfo{author}{\bibfnamefont{D.}~\bibnamefont{J\'erome}},
  \bibinfo{author}{\bibfnamefont{T.~M.} \bibnamefont{Rice}}, \bibnamefont{and}
  \bibinfo{author}{\bibfnamefont{W.}~\bibnamefont{Kohn}},
  \bibinfo{journal}{Phys. Rev.} \textbf{\bibinfo{volume}{158}},
  \bibinfo{pages}{462} (\bibinfo{year}{1967}),
  \urlprefix\url{https://link.aps.org/doi/10.1103/PhysRev.158.462}.

\bibitem[{\citenamefont{Halperin and Rice}(1968)}]{HalperinRice1968}
\bibinfo{author}{\bibfnamefont{B.}~\bibnamefont{Halperin}} \bibnamefont{and}
  \bibinfo{author}{\bibfnamefont{T.}~\bibnamefont{Rice}},
  \bibinfo{journal}{Solid State Physics} \textbf{\bibinfo{volume}{21}},
  \bibinfo{pages}{115 } (\bibinfo{year}{1968}),
  \urlprefix\url{http://www.sciencedirect.com/science/article/pii/S0081194708607407}.

\bibitem[{\citenamefont{Keldysh and Kozlov}(1968)}]{Keldysh-Kozlov_JETP1968}
\bibinfo{author}{\bibfnamefont{L.~V.} \bibnamefont{Keldysh}} \bibnamefont{and}
  \bibinfo{author}{\bibfnamefont{A.~N.} \bibnamefont{Kozlov}},
  \bibinfo{journal}{JETP} \textbf{\bibinfo{volume}{27}}, \bibinfo{pages}{521}
  (\bibinfo{year}{1968}).

\bibitem[{\citenamefont{Comte and Nozi\`eres}(1982)}]{Comte-Nozieres_1982}
\bibinfo{author}{\bibfnamefont{C.}~\bibnamefont{Comte}} \bibnamefont{and}
  \bibinfo{author}{\bibfnamefont{P.}~\bibnamefont{Nozi\`eres}},
  \bibinfo{journal}{J. Phys. France} \textbf{\bibinfo{volume}{43}},
  \bibinfo{pages}{1069Ð1081} (\bibinfo{year}{1982}).

\bibitem[{\citenamefont{Spielman et~al.}(2001)\citenamefont{Spielman,
  Eisenstein, Pfeiffer, and West}}]{Spielman_PhysRevLett.87.036803}
\bibinfo{author}{\bibfnamefont{I.~B.} \bibnamefont{Spielman}},
  \bibinfo{author}{\bibfnamefont{J.~P.} \bibnamefont{Eisenstein}},
  \bibinfo{author}{\bibfnamefont{L.~N.} \bibnamefont{Pfeiffer}},
  \bibnamefont{and} \bibinfo{author}{\bibfnamefont{K.~W.} \bibnamefont{West}},
  \bibinfo{journal}{Phys. Rev. Lett.} \textbf{\bibinfo{volume}{87}},
  \bibinfo{pages}{036803} (\bibinfo{year}{2001}),
  \urlprefix\url{https://link.aps.org/doi/10.1103/PhysRevLett.87.036803}.

\bibitem[{\citenamefont{Eisenstein and MacDonald}(2004)}]{Eisenstein2004}
\bibinfo{author}{\bibfnamefont{J.~P.} \bibnamefont{Eisenstein}}
  \bibnamefont{and} \bibinfo{author}{\bibfnamefont{A.~H.}
  \bibnamefont{MacDonald}}, \bibinfo{journal}{Nature}
  \textbf{\bibinfo{volume}{432}}, \bibinfo{pages}{691} (\bibinfo{year}{2004}),
  ISSN \bibinfo{issn}{1476-4687},
  \urlprefix\url{https://doi.org/10.1038/nature03081}.

\bibitem[{\citenamefont{Zhang and Joglekar}(2008)}]{Zhang_PhysRevB.77.233405}
\bibinfo{author}{\bibfnamefont{C.-H.} \bibnamefont{Zhang}} \bibnamefont{and}
  \bibinfo{author}{\bibfnamefont{Y.~N.} \bibnamefont{Joglekar}},
  \bibinfo{journal}{Phys. Rev. B} \textbf{\bibinfo{volume}{77}},
  \bibinfo{pages}{233405} (\bibinfo{year}{2008}),
  \urlprefix\url{https://link.aps.org/doi/10.1103/PhysRevB.77.233405}.

\bibitem[{\citenamefont{Min et~al.}(2008)\citenamefont{Min, Bistritzer, Su, and
  MacDonald}}]{Min_PhysRevB.78.121401}
\bibinfo{author}{\bibfnamefont{H.}~\bibnamefont{Min}},
  \bibinfo{author}{\bibfnamefont{R.}~\bibnamefont{Bistritzer}},
  \bibinfo{author}{\bibfnamefont{J.-J.} \bibnamefont{Su}}, \bibnamefont{and}
  \bibinfo{author}{\bibfnamefont{A.~H.} \bibnamefont{MacDonald}},
  \bibinfo{journal}{Phys. Rev. B} \textbf{\bibinfo{volume}{78}},
  \bibinfo{pages}{121401} (\bibinfo{year}{2008}),
  \urlprefix\url{https://link.aps.org/doi/10.1103/PhysRevB.78.121401}.

\bibitem[{\citenamefont{Lozovik and Sokolik}(2008)}]{Lozovik_JETP2008}
\bibinfo{author}{\bibfnamefont{Y.~E.} \bibnamefont{Lozovik}} \bibnamefont{and}
  \bibinfo{author}{\bibfnamefont{A.~A.} \bibnamefont{Sokolik}},
  \bibinfo{journal}{JETP Letters} \textbf{\bibinfo{volume}{87}},
  \bibinfo{pages}{55} (\bibinfo{year}{2008}), ISSN \bibinfo{issn}{1090-6487},
  \urlprefix\url{https://doi.org/10.1134/S002136400801013X}.

\bibitem[{\citenamefont{Kharitonov and
  Efetov}(2008)}]{Kharitonov_PhysRevB.78.241401}
\bibinfo{author}{\bibfnamefont{M.~Y.} \bibnamefont{Kharitonov}}
  \bibnamefont{and} \bibinfo{author}{\bibfnamefont{K.~B.}
  \bibnamefont{Efetov}}, \bibinfo{journal}{Phys. Rev. B}
  \textbf{\bibinfo{volume}{78}}, \bibinfo{pages}{241401}
  (\bibinfo{year}{2008}),
  \urlprefix\url{https://link.aps.org/doi/10.1103/PhysRevB.78.241401}.

\bibitem[{\citenamefont{Varsano et~al.}(2017)\citenamefont{Varsano, Sorella,
  Sangalli, Barborini, Corni, Molinari, and Rontani}}]{VarsanoNC2017}
\bibinfo{author}{\bibfnamefont{D.}~\bibnamefont{Varsano}},
  \bibinfo{author}{\bibfnamefont{S.}~\bibnamefont{Sorella}},
  \bibinfo{author}{\bibfnamefont{D.}~\bibnamefont{Sangalli}},
  \bibinfo{author}{\bibfnamefont{M.}~\bibnamefont{Barborini}},
  \bibinfo{author}{\bibfnamefont{S.}~\bibnamefont{Corni}},
  \bibinfo{author}{\bibfnamefont{E.}~\bibnamefont{Molinari}}, \bibnamefont{and}
  \bibinfo{author}{\bibfnamefont{M.}~\bibnamefont{Rontani}},
  \bibinfo{journal}{Nature Communications} \textbf{\bibinfo{volume}{8}},
  \bibinfo{pages}{1461} (\bibinfo{year}{2017}),
  \urlprefix\url{https://doi.org/10.1038/s41467-017-01660-8}.

\bibitem[{\citenamefont{Hellgren et~al.}(2018)\citenamefont{Hellgren, Baima,
  and Acheche}}]{Hellgren_PhysRevB.98.201103}
\bibinfo{author}{\bibfnamefont{M.}~\bibnamefont{Hellgren}},
  \bibinfo{author}{\bibfnamefont{J.}~\bibnamefont{Baima}}, \bibnamefont{and}
  \bibinfo{author}{\bibfnamefont{A.}~\bibnamefont{Acheche}},
  \bibinfo{journal}{Phys. Rev. B} \textbf{\bibinfo{volume}{98}},
  \bibinfo{pages}{201103} (\bibinfo{year}{2018}),
  \urlprefix\url{https://link.aps.org/doi/10.1103/PhysRevB.98.201103}.

\bibitem[{\citenamefont{Cercellier et~al.}(2007)\citenamefont{Cercellier,
  Monney, Clerc, Battaglia, Despont, Garnier, Beck, Aebi, Patthey, Berger
  et~al.}}]{Cercellier-PhysRevLett.99.146403}
\bibinfo{author}{\bibfnamefont{H.}~\bibnamefont{Cercellier}},
  \bibinfo{author}{\bibfnamefont{C.}~\bibnamefont{Monney}},
  \bibinfo{author}{\bibfnamefont{F.}~\bibnamefont{Clerc}},
  \bibinfo{author}{\bibfnamefont{C.}~\bibnamefont{Battaglia}},
  \bibinfo{author}{\bibfnamefont{L.}~\bibnamefont{Despont}},
  \bibinfo{author}{\bibfnamefont{M.~G.} \bibnamefont{Garnier}},
  \bibinfo{author}{\bibfnamefont{H.}~\bibnamefont{Beck}},
  \bibinfo{author}{\bibfnamefont{P.}~\bibnamefont{Aebi}},
  \bibinfo{author}{\bibfnamefont{L.}~\bibnamefont{Patthey}},
  \bibinfo{author}{\bibfnamefont{H.}~\bibnamefont{Berger}},
  \bibnamefont{et~al.}, \bibinfo{journal}{Phys. Rev. Lett.}
  \textbf{\bibinfo{volume}{99}}, \bibinfo{pages}{146403}
  (\bibinfo{year}{2007}),
  \urlprefix\url{https://link.aps.org/doi/10.1103/PhysRevLett.99.146403}.

\bibitem[{\citenamefont{Monney et~al.}(2012)\citenamefont{Monney, Monney, Aebi,
  and Beck}}]{Monney_NJP2012}
\bibinfo{author}{\bibfnamefont{C.}~\bibnamefont{Monney}},
  \bibinfo{author}{\bibfnamefont{G.}~\bibnamefont{Monney}},
  \bibinfo{author}{\bibfnamefont{P.}~\bibnamefont{Aebi}}, \bibnamefont{and}
  \bibinfo{author}{\bibfnamefont{H.}~\bibnamefont{Beck}}, \bibinfo{journal}{New
  Journal of Physics} \textbf{\bibinfo{volume}{14}}, \bibinfo{pages}{075026}
  (\bibinfo{year}{2012}),
  \urlprefix\url{https://doi.org/10.1088%2F1367-2630%2F14%2F7%2F075026}.

\bibitem[{\citenamefont{Zenker et~al.}(2013)\citenamefont{Zenker, Fehske, Beck,
  Monney, and Bishop}}]{Zenker_PhysRevB.88.075138}
\bibinfo{author}{\bibfnamefont{B.}~\bibnamefont{Zenker}},
  \bibinfo{author}{\bibfnamefont{H.}~\bibnamefont{Fehske}},
  \bibinfo{author}{\bibfnamefont{H.}~\bibnamefont{Beck}},
  \bibinfo{author}{\bibfnamefont{C.}~\bibnamefont{Monney}}, \bibnamefont{and}
  \bibinfo{author}{\bibfnamefont{A.~R.} \bibnamefont{Bishop}},
  \bibinfo{journal}{Phys. Rev. B} \textbf{\bibinfo{volume}{88}},
  \bibinfo{pages}{075138} (\bibinfo{year}{2013}),
  \urlprefix\url{https://link.aps.org/doi/10.1103/PhysRevB.88.075138}.

\bibitem[{\citenamefont{Zenker et~al.}(2014)\citenamefont{Zenker, Fehske, and
  Beck}}]{Zenker_PhysRevB.90.195118}
\bibinfo{author}{\bibfnamefont{B.}~\bibnamefont{Zenker}},
  \bibinfo{author}{\bibfnamefont{H.}~\bibnamefont{Fehske}}, \bibnamefont{and}
  \bibinfo{author}{\bibfnamefont{H.}~\bibnamefont{Beck}},
  \bibinfo{journal}{Phys. Rev. B} \textbf{\bibinfo{volume}{90}},
  \bibinfo{pages}{195118} (\bibinfo{year}{2014}),
  \urlprefix\url{https://link.aps.org/doi/10.1103/PhysRevB.90.195118}.

\bibitem[{\citenamefont{Kogar et~al.}(2017)\citenamefont{Kogar, Rak, Vig,
  Husain, Flicker, Joe, Venema, MacDougall, Chiang, Fradkin
  et~al.}}]{Kogar2017}
\bibinfo{author}{\bibfnamefont{A.}~\bibnamefont{Kogar}},
  \bibinfo{author}{\bibfnamefont{M.~S.} \bibnamefont{Rak}},
  \bibinfo{author}{\bibfnamefont{S.}~\bibnamefont{Vig}},
  \bibinfo{author}{\bibfnamefont{A.~A.} \bibnamefont{Husain}},
  \bibinfo{author}{\bibfnamefont{F.}~\bibnamefont{Flicker}},
  \bibinfo{author}{\bibfnamefont{Y.~I.} \bibnamefont{Joe}},
  \bibinfo{author}{\bibfnamefont{L.}~\bibnamefont{Venema}},
  \bibinfo{author}{\bibfnamefont{G.~J.} \bibnamefont{MacDougall}},
  \bibinfo{author}{\bibfnamefont{T.~C.} \bibnamefont{Chiang}},
  \bibinfo{author}{\bibfnamefont{E.}~\bibnamefont{Fradkin}},
  \bibnamefont{et~al.}, \bibinfo{journal}{Science}
  \textbf{\bibinfo{volume}{358}}, \bibinfo{pages}{1314} (\bibinfo{year}{2017}),
  \urlprefix\url{https://science.sciencemag.org/content/358/6368/1314}.

\bibitem[{\citenamefont{Gole\ifmmode~\check{z}\else \v{z}\fi{}
  et~al.}(2016)\citenamefont{Gole\ifmmode~\check{z}\else \v{z}\fi{}, Werner,
  and Eckstein}}]{GolezPRB2016}
\bibinfo{author}{\bibfnamefont{D.}~\bibnamefont{Gole\ifmmode~\check{z}\else
  \v{z}\fi{}}}, \bibinfo{author}{\bibfnamefont{P.}~\bibnamefont{Werner}},
  \bibnamefont{and} \bibinfo{author}{\bibfnamefont{M.}~\bibnamefont{Eckstein}},
  \bibinfo{journal}{Phys. Rev. B} \textbf{\bibinfo{volume}{94}},
  \bibinfo{pages}{035121} (\bibinfo{year}{2016}),
  \urlprefix\url{https://link.aps.org/doi/10.1103/PhysRevB.94.035121}.

\bibitem[{\citenamefont{Murakami et~al.}(2017)\citenamefont{Murakami,
  Gole\ifmmode~\check{z}\else \v{z}\fi{}, Eckstein, and
  Werner}}]{MurakamiPRL2017}
\bibinfo{author}{\bibfnamefont{Y.}~\bibnamefont{Murakami}},
  \bibinfo{author}{\bibfnamefont{D.}~\bibnamefont{Gole\ifmmode~\check{z}\else
  \v{z}\fi{}}}, \bibinfo{author}{\bibfnamefont{M.}~\bibnamefont{Eckstein}},
  \bibnamefont{and} \bibinfo{author}{\bibfnamefont{P.}~\bibnamefont{Werner}},
  \bibinfo{journal}{Phys. Rev. Lett.} \textbf{\bibinfo{volume}{119}},
  \bibinfo{pages}{247601} (\bibinfo{year}{2017}),
  \urlprefix\url{https://link.aps.org/doi/10.1103/PhysRevLett.119.247601}.

\bibitem[{\citenamefont{Mor et~al.}(2017)\citenamefont{Mor, Herzog,
  Gole\ifmmode~\check{z}\else \v{z}\fi{}, Werner, Eckstein, Katayama, Nohara,
  Takagi, Mizokawa, Monney et~al.}}]{MorPRL2017}
\bibinfo{author}{\bibfnamefont{S.}~\bibnamefont{Mor}},
  \bibinfo{author}{\bibfnamefont{M.}~\bibnamefont{Herzog}},
  \bibinfo{author}{\bibfnamefont{D.}~\bibnamefont{Gole\ifmmode~\check{z}\else
  \v{z}\fi{}}}, \bibinfo{author}{\bibfnamefont{P.}~\bibnamefont{Werner}},
  \bibinfo{author}{\bibfnamefont{M.}~\bibnamefont{Eckstein}},
  \bibinfo{author}{\bibfnamefont{N.}~\bibnamefont{Katayama}},
  \bibinfo{author}{\bibfnamefont{M.}~\bibnamefont{Nohara}},
  \bibinfo{author}{\bibfnamefont{H.}~\bibnamefont{Takagi}},
  \bibinfo{author}{\bibfnamefont{T.}~\bibnamefont{Mizokawa}},
  \bibinfo{author}{\bibfnamefont{C.}~\bibnamefont{Monney}},
  \bibnamefont{et~al.}, \bibinfo{journal}{Phys. Rev. Lett.}
  \textbf{\bibinfo{volume}{119}}, \bibinfo{pages}{086401}
  (\bibinfo{year}{2017}),
  \urlprefix\url{https://link.aps.org/doi/10.1103/PhysRevLett.119.086401}.

\bibitem[{\citenamefont{Werdehausen
  et~al.}(2018{\natexlab{a}})\citenamefont{Werdehausen, Takayama, H{\"o}ppner,
  Albrecht, Rost, Lu, Manske, Takagi, and Kaiser}}]{Werdehauseneaap8652}
\bibinfo{author}{\bibfnamefont{D.}~\bibnamefont{Werdehausen}},
  \bibinfo{author}{\bibfnamefont{T.}~\bibnamefont{Takayama}},
  \bibinfo{author}{\bibfnamefont{M.}~\bibnamefont{H{\"o}ppner}},
  \bibinfo{author}{\bibfnamefont{G.}~\bibnamefont{Albrecht}},
  \bibinfo{author}{\bibfnamefont{A.~W.} \bibnamefont{Rost}},
  \bibinfo{author}{\bibfnamefont{Y.}~\bibnamefont{Lu}},
  \bibinfo{author}{\bibfnamefont{D.}~\bibnamefont{Manske}},
  \bibinfo{author}{\bibfnamefont{H.}~\bibnamefont{Takagi}}, \bibnamefont{and}
  \bibinfo{author}{\bibfnamefont{S.}~\bibnamefont{Kaiser}},
  \bibinfo{journal}{Science Advances} \textbf{\bibinfo{volume}{4}}
  (\bibinfo{year}{2018}{\natexlab{a}}),
  \urlprefix\url{https://advances.sciencemag.org/content/4/3/eaap8652}.

\bibitem[{\citenamefont{Werdehausen
  et~al.}(2018{\natexlab{b}})\citenamefont{Werdehausen, Takayama, Albrecht, Lu,
  Takagi, and Kaiser}}]{Werdehausen_2018}
\bibinfo{author}{\bibfnamefont{D.}~\bibnamefont{Werdehausen}},
  \bibinfo{author}{\bibfnamefont{T.}~\bibnamefont{Takayama}},
  \bibinfo{author}{\bibfnamefont{G.}~\bibnamefont{Albrecht}},
  \bibinfo{author}{\bibfnamefont{Y.}~\bibnamefont{Lu}},
  \bibinfo{author}{\bibfnamefont{H.}~\bibnamefont{Takagi}}, \bibnamefont{and}
  \bibinfo{author}{\bibfnamefont{S.}~\bibnamefont{Kaiser}},
  \bibinfo{journal}{Journal of Physics: Condensed Matter}
  \textbf{\bibinfo{volume}{30}}, \bibinfo{pages}{305602}
  (\bibinfo{year}{2018}{\natexlab{b}}),
  \urlprefix\url{https://doi.org/10.1088%2F1361-648x%2Faacd76}.

\bibitem[{\citenamefont{Tanabe et~al.}(2018)\citenamefont{Tanabe, Sugimoto, and
  Ohta}}]{Tanabe_PhysRevB.98.235127}
\bibinfo{author}{\bibfnamefont{T.}~\bibnamefont{Tanabe}},
  \bibinfo{author}{\bibfnamefont{K.}~\bibnamefont{Sugimoto}}, \bibnamefont{and}
  \bibinfo{author}{\bibfnamefont{Y.}~\bibnamefont{Ohta}},
  \bibinfo{journal}{Phys. Rev. B} \textbf{\bibinfo{volume}{98}},
  \bibinfo{pages}{235127} (\bibinfo{year}{2018}),
  \urlprefix\url{https://link.aps.org/doi/10.1103/PhysRevB.98.235127}.

\bibitem[{\citenamefont{Tuovinen et~al.}(2019)\citenamefont{Tuovinen, Golez,
  SchŸler, Werner, Eckstein, and Sentef}}]{Tuovinen_pssb2019}
\bibinfo{author}{\bibfnamefont{R.}~\bibnamefont{Tuovinen}},
  \bibinfo{author}{\bibfnamefont{D.}~\bibnamefont{Golez}},
  \bibinfo{author}{\bibfnamefont{M.}~\bibnamefont{SchŸler}},
  \bibinfo{author}{\bibfnamefont{P.}~\bibnamefont{Werner}},
  \bibinfo{author}{\bibfnamefont{M.}~\bibnamefont{Eckstein}}, \bibnamefont{and}
  \bibinfo{author}{\bibfnamefont{M.~A.} \bibnamefont{Sentef}},
  \bibinfo{journal}{physica status solidi (b)} p. \bibinfo{pages}{1800469}
  (\bibinfo{year}{2019}),
  \urlprefix\url{https://onlinelibrary.wiley.com/doi/abs/10.1002/pssb.201800469}.

\bibitem[{\citenamefont{Wakisaka et~al.}(2009)\citenamefont{Wakisaka, Sudayama,
  Takubo, Mizokawa, Arita, Namatame, Taniguchi, Katayama, Nohara, and
  Takagi}}]{Wakisaka_PhysRevLett.103.026402}
\bibinfo{author}{\bibfnamefont{Y.}~\bibnamefont{Wakisaka}},
  \bibinfo{author}{\bibfnamefont{T.}~\bibnamefont{Sudayama}},
  \bibinfo{author}{\bibfnamefont{K.}~\bibnamefont{Takubo}},
  \bibinfo{author}{\bibfnamefont{T.}~\bibnamefont{Mizokawa}},
  \bibinfo{author}{\bibfnamefont{M.}~\bibnamefont{Arita}},
  \bibinfo{author}{\bibfnamefont{H.}~\bibnamefont{Namatame}},
  \bibinfo{author}{\bibfnamefont{M.}~\bibnamefont{Taniguchi}},
  \bibinfo{author}{\bibfnamefont{N.}~\bibnamefont{Katayama}},
  \bibinfo{author}{\bibfnamefont{M.}~\bibnamefont{Nohara}}, \bibnamefont{and}
  \bibinfo{author}{\bibfnamefont{H.}~\bibnamefont{Takagi}},
  \bibinfo{journal}{Phys. Rev. Lett.} \textbf{\bibinfo{volume}{103}},
  \bibinfo{pages}{026402} (\bibinfo{year}{2009}),
  \urlprefix\url{https://link.aps.org/doi/10.1103/PhysRevLett.103.026402}.

\bibitem[{\citenamefont{Seki et~al.}(2014)\citenamefont{Seki, Wakisaka, Kaneko,
  Toriyama, Konishi, Sudayama, Saini, Arita, Namatame, Taniguchi
  et~al.}}]{Seki_PhysRevB.90.155116}
\bibinfo{author}{\bibfnamefont{K.}~\bibnamefont{Seki}},
  \bibinfo{author}{\bibfnamefont{Y.}~\bibnamefont{Wakisaka}},
  \bibinfo{author}{\bibfnamefont{T.}~\bibnamefont{Kaneko}},
  \bibinfo{author}{\bibfnamefont{T.}~\bibnamefont{Toriyama}},
  \bibinfo{author}{\bibfnamefont{T.}~\bibnamefont{Konishi}},
  \bibinfo{author}{\bibfnamefont{T.}~\bibnamefont{Sudayama}},
  \bibinfo{author}{\bibfnamefont{N.~L.} \bibnamefont{Saini}},
  \bibinfo{author}{\bibfnamefont{M.}~\bibnamefont{Arita}},
  \bibinfo{author}{\bibfnamefont{H.}~\bibnamefont{Namatame}},
  \bibinfo{author}{\bibfnamefont{M.}~\bibnamefont{Taniguchi}},
  \bibnamefont{et~al.}, \bibinfo{journal}{Phys. Rev. B}
  \textbf{\bibinfo{volume}{90}}, \bibinfo{pages}{155116}
  (\bibinfo{year}{2014}),
  \urlprefix\url{https://link.aps.org/doi/10.1103/PhysRevB.90.155116}.

\bibitem[{\citenamefont{Triola et~al.}(2017)\citenamefont{Triola, Pertsova,
  Markiewicz, and Balatsky}}]{TriolaPRB2017}
\bibinfo{author}{\bibfnamefont{C.}~\bibnamefont{Triola}},
  \bibinfo{author}{\bibfnamefont{A.}~\bibnamefont{Pertsova}},
  \bibinfo{author}{\bibfnamefont{R.~S.} \bibnamefont{Markiewicz}},
  \bibnamefont{and} \bibinfo{author}{\bibfnamefont{A.~V.}
  \bibnamefont{Balatsky}}, \bibinfo{journal}{Phys. Rev. B}
  \textbf{\bibinfo{volume}{95}}, \bibinfo{pages}{205410}
  (\bibinfo{year}{2017}),
  \urlprefix\url{https://link.aps.org/doi/10.1103/PhysRevB.95.205410}.

\bibitem[{\citenamefont{Pertsova and Balatsky}(2018)}]{PertsovaPRB2018}
\bibinfo{author}{\bibfnamefont{A.}~\bibnamefont{Pertsova}} \bibnamefont{and}
  \bibinfo{author}{\bibfnamefont{A.~V.} \bibnamefont{Balatsky}},
  \bibinfo{journal}{Phys. Rev. B} \textbf{\bibinfo{volume}{97}},
  \bibinfo{pages}{075109} (\bibinfo{year}{2018}),
  \urlprefix\url{https://link.aps.org/doi/10.1103/PhysRevB.97.075109}.

\bibitem[{\citenamefont{Szyma\ifmmode~\acute{n}\else \'{n}\fi{}ska
  et~al.}(2006)\citenamefont{Szyma\ifmmode~\acute{n}\else \'{n}\fi{}ska,
  Keeling, and Littlewood}}]{SzymaPRL2006}
\bibinfo{author}{\bibfnamefont{M.~H.} \bibnamefont{Szyma\ifmmode~\acute{n}\else
  \'{n}\fi{}ska}}, \bibinfo{author}{\bibfnamefont{J.}~\bibnamefont{Keeling}},
  \bibnamefont{and} \bibinfo{author}{\bibfnamefont{P.~B.}
  \bibnamefont{Littlewood}}, \bibinfo{journal}{Phys. Rev. Lett.}
  \textbf{\bibinfo{volume}{96}}, \bibinfo{pages}{230602}
  (\bibinfo{year}{2006}),
  \urlprefix\url{https://link.aps.org/doi/10.1103/PhysRevLett.96.230602}.

\bibitem[{\citenamefont{Hanai et~al.}(2016)\citenamefont{Hanai, Littlewood, and
  Ohashi}}]{Hanai2016}
\bibinfo{author}{\bibfnamefont{R.}~\bibnamefont{Hanai}},
  \bibinfo{author}{\bibfnamefont{P.~B.} \bibnamefont{Littlewood}},
  \bibnamefont{and} \bibinfo{author}{\bibfnamefont{Y.}~\bibnamefont{Ohashi}},
  \bibinfo{journal}{Journal of Low Temperature Physics}
  \textbf{\bibinfo{volume}{183}}, \bibinfo{pages}{127} (\bibinfo{year}{2016}),
  \urlprefix\url{https://doi.org/10.1007/s10909-016-1552-6}.

\bibitem[{\citenamefont{Hanai et~al.}(2017)\citenamefont{Hanai, Littlewood, and
  Ohashi}}]{HanaiPRB2017}
\bibinfo{author}{\bibfnamefont{R.}~\bibnamefont{Hanai}},
  \bibinfo{author}{\bibfnamefont{P.~B.} \bibnamefont{Littlewood}},
  \bibnamefont{and} \bibinfo{author}{\bibfnamefont{Y.}~\bibnamefont{Ohashi}},
  \bibinfo{journal}{Phys. Rev. B} \textbf{\bibinfo{volume}{96}},
  \bibinfo{pages}{125206} (\bibinfo{year}{2017}),
  \urlprefix\url{https://link.aps.org/doi/10.1103/PhysRevB.96.125206}.

\bibitem[{\citenamefont{Hanai et~al.}(2018)\citenamefont{Hanai, Littlewood, and
  Ohashi}}]{Hanai2018}
\bibinfo{author}{\bibfnamefont{R.}~\bibnamefont{Hanai}},
  \bibinfo{author}{\bibfnamefont{P.~B.} \bibnamefont{Littlewood}},
  \bibnamefont{and} \bibinfo{author}{\bibfnamefont{Y.}~\bibnamefont{Ohashi}},
  \bibinfo{journal}{Phys. Rev. B} \textbf{\bibinfo{volume}{97}},
  \bibinfo{pages}{245302} (\bibinfo{year}{2018}),
  \urlprefix\url{https://link.aps.org/doi/10.1103/PhysRevB.97.245302}.

\bibitem[{\citenamefont{Becker et~al.}(2019)\citenamefont{Becker, Fehske, and
  Phan}}]{Becker_PhysRevB.99.035304}
\bibinfo{author}{\bibfnamefont{K.~W.} \bibnamefont{Becker}},
  \bibinfo{author}{\bibfnamefont{H.}~\bibnamefont{Fehske}}, \bibnamefont{and}
  \bibinfo{author}{\bibfnamefont{V.-N.} \bibnamefont{Phan}},
  \bibinfo{journal}{Phys. Rev. B} \textbf{\bibinfo{volume}{99}},
  \bibinfo{pages}{035304} (\bibinfo{year}{2019}),
  \urlprefix\url{https://link.aps.org/doi/10.1103/PhysRevB.99.035304}.

\bibitem[{\citenamefont{Yamaguchi et~al.}(2012)\citenamefont{Yamaguchi, Kamide,
  Ogawa, and Yamamoto}}]{Yamaguchi_NJP2012}
\bibinfo{author}{\bibfnamefont{M.}~\bibnamefont{Yamaguchi}},
  \bibinfo{author}{\bibfnamefont{K.}~\bibnamefont{Kamide}},
  \bibinfo{author}{\bibfnamefont{T.}~\bibnamefont{Ogawa}}, \bibnamefont{and}
  \bibinfo{author}{\bibfnamefont{Y.}~\bibnamefont{Yamamoto}},
  \bibinfo{journal}{New Journal of Physics} \textbf{\bibinfo{volume}{14}},
  \bibinfo{pages}{065001} (\bibinfo{year}{2012}),
  \urlprefix\url{https://doi.org/10.1088%2F1367-2630%2F14%2F6%2F065001}.

\bibitem[{\citenamefont{Yamaguchi et~al.}(2013)\citenamefont{Yamaguchi, Kamide,
  Nii, Ogawa, and Yamamoto}}]{Yamaguchi_PhysRevLett.111.026404}
\bibinfo{author}{\bibfnamefont{M.}~\bibnamefont{Yamaguchi}},
  \bibinfo{author}{\bibfnamefont{K.}~\bibnamefont{Kamide}},
  \bibinfo{author}{\bibfnamefont{R.}~\bibnamefont{Nii}},
  \bibinfo{author}{\bibfnamefont{T.}~\bibnamefont{Ogawa}}, \bibnamefont{and}
  \bibinfo{author}{\bibfnamefont{Y.}~\bibnamefont{Yamamoto}},
  \bibinfo{journal}{Phys. Rev. Lett.} \textbf{\bibinfo{volume}{111}},
  \bibinfo{pages}{026404} (\bibinfo{year}{2013}),
  \urlprefix\url{https://link.aps.org/doi/10.1103/PhysRevLett.111.026404}.

\bibitem[{\citenamefont{Schmitt-Rink et~al.}(1988)\citenamefont{Schmitt-Rink,
  Chemla, and Haug}}]{Schmitt-Rink_PhysRevB.37.941}
\bibinfo{author}{\bibfnamefont{S.}~\bibnamefont{Schmitt-Rink}},
  \bibinfo{author}{\bibfnamefont{D.~S.} \bibnamefont{Chemla}},
  \bibnamefont{and} \bibinfo{author}{\bibfnamefont{H.}~\bibnamefont{Haug}},
  \bibinfo{journal}{Phys. Rev. B} \textbf{\bibinfo{volume}{37}},
  \bibinfo{pages}{941} (\bibinfo{year}{1988}),
  \urlprefix\url{https://link.aps.org/doi/10.1103/PhysRevB.37.941}.

\bibitem[{\citenamefont{Haug}(1985)}]{HAUG_1985}
\bibinfo{author}{\bibfnamefont{H.}~\bibnamefont{Haug}},
  \bibinfo{journal}{Journal of Luminescence} \textbf{\bibinfo{volume}{30}},
  \bibinfo{pages}{171 } (\bibinfo{year}{1985}),
  \urlprefix\url{http://www.sciencedirect.com/science/article/pii/0022231385900511}.

\bibitem[{\citenamefont{Glutsch and
  Zimmermann}(1992)}]{Glutsch_PhysRevB.45.5857}
\bibinfo{author}{\bibfnamefont{S.}~\bibnamefont{Glutsch}} \bibnamefont{and}
  \bibinfo{author}{\bibfnamefont{R.}~\bibnamefont{Zimmermann}},
  \bibinfo{journal}{Phys. Rev. B} \textbf{\bibinfo{volume}{45}},
  \bibinfo{pages}{5857} (\bibinfo{year}{1992}),
  \urlprefix\url{https://link.aps.org/doi/10.1103/PhysRevB.45.5857}.

\bibitem[{\citenamefont{Chu and Chang}(1996)}]{Chu_PhysRevB.54.5020}
\bibinfo{author}{\bibfnamefont{H.}~\bibnamefont{Chu}} \bibnamefont{and}
  \bibinfo{author}{\bibfnamefont{Y.~C.} \bibnamefont{Chang}},
  \bibinfo{journal}{Phys. Rev. B} \textbf{\bibinfo{volume}{54}},
  \bibinfo{pages}{5020} (\bibinfo{year}{1996}),
  \urlprefix\url{https://link.aps.org/doi/10.1103/PhysRevB.54.5020}.

\bibitem[{\citenamefont{{\"O}streich and
  Sch{\"o}nhammer}(1993)}]{Ostreich_1993}
\bibinfo{author}{\bibfnamefont{T.}~\bibnamefont{{\"O}streich}}
  \bibnamefont{and}
  \bibinfo{author}{\bibfnamefont{K.}~\bibnamefont{Sch{\"o}nhammer}},
  \bibinfo{journal}{Zeitschrift f{\"u}r Physik B Condensed Matter}
  \textbf{\bibinfo{volume}{91}}, \bibinfo{pages}{189} (\bibinfo{year}{1993}),
  \urlprefix\url{https://doi.org/10.1007/BF01315235}.

\bibitem[{\citenamefont{Jun et~al.}(2017)\citenamefont{Jun, Mervin, Yuan, and
  Xiang}}]{Xiao2017}
\bibinfo{author}{\bibfnamefont{X.}~\bibnamefont{Jun}},
  \bibinfo{author}{\bibfnamefont{Z.}~\bibnamefont{Mervin}},
  \bibinfo{author}{\bibfnamefont{W.}~\bibnamefont{Yuan}}, \bibnamefont{and}
  \bibinfo{author}{\bibfnamefont{Z.}~\bibnamefont{Xiang}},
  \bibinfo{journal}{Nanoph.} \textbf{\bibinfo{volume}{6}},
  \bibinfo{pages}{1309} (\bibinfo{year}{2017}),
  \urlprefix\url{https://www.degruyter.com/view/j/nanoph.2017.6.issue-6/nanoph-2016-0160/nanoph-2016-0160.xml}.

\bibitem[{\citenamefont{Hannewald et~al.}(2000)\citenamefont{Hannewald,
  Glutsch, and Bechstedt}}]{Hannewald-Bechstedt_2000}
\bibinfo{author}{\bibfnamefont{K.}~\bibnamefont{Hannewald}},
  \bibinfo{author}{\bibfnamefont{S.}~\bibnamefont{Glutsch}}, \bibnamefont{and}
  \bibinfo{author}{\bibfnamefont{F.}~\bibnamefont{Bechstedt}},
  \bibinfo{journal}{Journal of Physics: Condensed Matter}
  \textbf{\bibinfo{volume}{13}}, \bibinfo{pages}{275} (\bibinfo{year}{2000}),
  \urlprefix\url{https://doi.org/10.1088%2F0953-8984%2F13%2F2%2F305}.

\bibitem[{\citenamefont{Marini et~al.}(2009)\citenamefont{Marini, Hogan,
  Gr\"uning, and Varsano}}]{MARINI20091392}
\bibinfo{author}{\bibfnamefont{A.}~\bibnamefont{Marini}},
  \bibinfo{author}{\bibfnamefont{C.}~\bibnamefont{Hogan}},
  \bibinfo{author}{\bibfnamefont{M.}~\bibnamefont{Gr\"uning}},
  \bibnamefont{and} \bibinfo{author}{\bibfnamefont{D.}~\bibnamefont{Varsano}},
  \bibinfo{journal}{Computer Physics Communications}
  \textbf{\bibinfo{volume}{180}}, \bibinfo{pages}{1392 }
  (\bibinfo{year}{2009}),
  \urlprefix\url{http://www.sciencedirect.com/science/article/pii/S0010465509000472}.

\bibitem[{\citenamefont{Eagles}(1969)}]{Eagles_PhysRev.186.456}
\bibinfo{author}{\bibfnamefont{D.~M.} \bibnamefont{Eagles}},
  \bibinfo{journal}{Phys. Rev.} \textbf{\bibinfo{volume}{186}},
  \bibinfo{pages}{456} (\bibinfo{year}{1969}),
  \urlprefix\url{https://link.aps.org/doi/10.1103/PhysRev.186.456}.

\bibitem[{\citenamefont{Strinati et~al.}(2018)\citenamefont{Strinati, Pieri,
  Ršpke, Schuck, and Urban}}]{Strinati-2018}
\bibinfo{author}{\bibfnamefont{G.~C.} \bibnamefont{Strinati}},
  \bibinfo{author}{\bibfnamefont{P.}~\bibnamefont{Pieri}},
  \bibinfo{author}{\bibfnamefont{G.}~\bibnamefont{Ršpke}},
  \bibinfo{author}{\bibfnamefont{P.}~\bibnamefont{Schuck}}, \bibnamefont{and}
  \bibinfo{author}{\bibfnamefont{M.}~\bibnamefont{Urban}},
  \bibinfo{journal}{Physics Reports} \textbf{\bibinfo{volume}{738}},
  \bibinfo{pages}{1 } (\bibinfo{year}{2018}),
  \urlprefix\url{http://www.sciencedirect.com/science/article/pii/S0370157318300267}.

\bibitem[{\citenamefont{Yang et~al.}(2012)\citenamefont{Yang, Li, and
  Ullrich}}]{YLU.2012}
\bibinfo{author}{\bibfnamefont{Z.-h.} \bibnamefont{Yang}},
  \bibinfo{author}{\bibfnamefont{Y.}~\bibnamefont{Li}}, \bibnamefont{and}
  \bibinfo{author}{\bibfnamefont{C.~A.} \bibnamefont{Ullrich}},
  \bibinfo{journal}{The Journal of Chemical Physics}
  \textbf{\bibinfo{volume}{137}}, \bibinfo{pages}{014513}
  (\bibinfo{year}{2012}), \urlprefix\url{https://doi.org/10.1063/1.4730031}.

\bibitem[{\citenamefont{Stefanucci and van Leeuwen}(2013)}]{svl-book}
\bibinfo{author}{\bibfnamefont{G.}~\bibnamefont{Stefanucci}} \bibnamefont{and}
  \bibinfo{author}{\bibfnamefont{R.}~\bibnamefont{van Leeuwen}},
  \emph{\bibinfo{title}{Nonequilibrium Many-Body Theory of Quantum Systems: A
  Modern Introduction}} (\bibinfo{publisher}{Cambridge University Press},
  \bibinfo{address}{Cambridge}, \bibinfo{year}{2013}).

\bibitem[{\citenamefont{Konstantinov and Perel'}(1961)}]{kp.1961}
\bibinfo{author}{\bibfnamefont{O.~V.} \bibnamefont{Konstantinov}}
  \bibnamefont{and} \bibinfo{author}{\bibfnamefont{V.~I.}
  \bibnamefont{Perel'}}, \bibinfo{journal}{Sov. Phys. JETP}
  \textbf{\bibinfo{volume}{12}}, \bibinfo{pages}{142} (\bibinfo{year}{1961}).

\bibitem[{\citenamefont{Rontani and Sham}(2009)}]{Rontani_PhysRevB.80.075309}
\bibinfo{author}{\bibfnamefont{M.}~\bibnamefont{Rontani}} \bibnamefont{and}
  \bibinfo{author}{\bibfnamefont{L.~J.} \bibnamefont{Sham}},
  \bibinfo{journal}{Phys. Rev. B} \textbf{\bibinfo{volume}{80}},
  \bibinfo{pages}{075309} (\bibinfo{year}{2009}),
  \urlprefix\url{https://link.aps.org/doi/10.1103/PhysRevB.80.075309}.

\bibitem[{\citenamefont{Joglekar et~al.}(2005)\citenamefont{Joglekar, Balatsky,
  and Lilly}}]{Joglekar_PhysRevB.72.205313}
\bibinfo{author}{\bibfnamefont{Y.~N.} \bibnamefont{Joglekar}},
  \bibinfo{author}{\bibfnamefont{A.~V.} \bibnamefont{Balatsky}},
  \bibnamefont{and} \bibinfo{author}{\bibfnamefont{M.~P.} \bibnamefont{Lilly}},
  \bibinfo{journal}{Phys. Rev. B} \textbf{\bibinfo{volume}{72}},
  \bibinfo{pages}{205313} (\bibinfo{year}{2005}),
  \urlprefix\url{https://link.aps.org/doi/10.1103/PhysRevB.72.205313}.

\bibitem[{\citenamefont{Hsu and Su}(2015)}]{Hsu2015}
\bibinfo{author}{\bibfnamefont{Y.-F.} \bibnamefont{Hsu}} \bibnamefont{and}
  \bibinfo{author}{\bibfnamefont{J.-J.} \bibnamefont{Su}},
  \bibinfo{journal}{Scientific Reports} \textbf{\bibinfo{volume}{5}},
  \bibinfo{pages}{15796} (\bibinfo{year}{2015}),
  \urlprefix\url{https://doi.org/10.1038/srep15796}.

\bibitem[{\citenamefont{Apinyan and Kope{\'{c}}}(2019)}]{Apinyan2019}
\bibinfo{author}{\bibfnamefont{V.}~\bibnamefont{Apinyan}} \bibnamefont{and}
  \bibinfo{author}{\bibfnamefont{T.~K.} \bibnamefont{Kope{\'{c}}}},
  \bibinfo{journal}{Journal of Low Temperature Physics}
  \textbf{\bibinfo{volume}{194}}, \bibinfo{pages}{325} (\bibinfo{year}{2019}),
  ISSN \bibinfo{issn}{1573-7357},
  \urlprefix\url{https://doi.org/10.1007/s10909-018-2107-9}.

\bibitem[{\citenamefont{Rustagi and Kemper}(2018)}]{RustagiKemper2018}
\bibinfo{author}{\bibfnamefont{A.}~\bibnamefont{Rustagi}} \bibnamefont{and}
  \bibinfo{author}{\bibfnamefont{A.~F.} \bibnamefont{Kemper}},
  \bibinfo{journal}{Phys. Rev. B} \textbf{\bibinfo{volume}{97}},
  \bibinfo{pages}{235310} (\bibinfo{year}{2018}),
  \urlprefix\url{https://link.aps.org/doi/10.1103/PhysRevB.97.235310}.

\bibitem[{\citenamefont{Nozi{\`e}res and Schmitt-Rink}(1985)}]{Nozieres1985}
\bibinfo{author}{\bibfnamefont{P.}~\bibnamefont{Nozi{\`e}res}}
  \bibnamefont{and}
  \bibinfo{author}{\bibfnamefont{S.}~\bibnamefont{Schmitt-Rink}},
  \bibinfo{journal}{Journal of Low Temperature Physics}
  \textbf{\bibinfo{volume}{59}}, \bibinfo{pages}{195} (\bibinfo{year}{1985}),
  \urlprefix\url{https://doi.org/10.1007/BF00683774}.

\bibitem[{\citenamefont{Roessler and Walker}(1967)}]{Roessler:67}
\bibinfo{author}{\bibfnamefont{D.~M.} \bibnamefont{Roessler}} \bibnamefont{and}
  \bibinfo{author}{\bibfnamefont{W.~C.} \bibnamefont{Walker}},
  \bibinfo{journal}{J. Opt. Soc. Am.} \textbf{\bibinfo{volume}{57}},
  \bibinfo{pages}{835} (\bibinfo{year}{1967}),
  \urlprefix\url{http://www.osapublishing.org/abstract.cfm?URI=josa-57-6-835}.

\bibitem[{\citenamefont{Giannozzi et~al.}(2009)\citenamefont{Giannozzi, Baroni,
  Bonini, Calandra, Car, Cavazzoni, Ceresoli, Chiarotti, Cococcioni, Dabo
  et~al.}}]{QuantumEspresso}
\bibinfo{author}{\bibfnamefont{P.}~\bibnamefont{Giannozzi}},
  \bibinfo{author}{\bibfnamefont{S.}~\bibnamefont{Baroni}},
  \bibinfo{author}{\bibfnamefont{N.}~\bibnamefont{Bonini}},
  \bibinfo{author}{\bibfnamefont{M.}~\bibnamefont{Calandra}},
  \bibinfo{author}{\bibfnamefont{R.}~\bibnamefont{Car}},
  \bibinfo{author}{\bibfnamefont{C.}~\bibnamefont{Cavazzoni}},
  \bibinfo{author}{\bibfnamefont{D.}~\bibnamefont{Ceresoli}},
  \bibinfo{author}{\bibfnamefont{G.~L.} \bibnamefont{Chiarotti}},
  \bibinfo{author}{\bibfnamefont{M.}~\bibnamefont{Cococcioni}},
  \bibinfo{author}{\bibfnamefont{I.}~\bibnamefont{Dabo}}, \bibnamefont{et~al.},
  \bibinfo{journal}{Journal of Physics: Condensed Matter}
  \textbf{\bibinfo{volume}{21}}, \bibinfo{pages}{395502}
  (\bibinfo{year}{2009}),
  \urlprefix\url{http://stacks.iop.org/0953-8984/21/i=39/a=395502}.

\bibitem[{\citenamefont{Wang et~al.}(2003)\citenamefont{Wang, Rohlfing,
  Kr\"uger, and Pollmann}}]{Wang2003}
\bibinfo{author}{\bibfnamefont{N.-P.} \bibnamefont{Wang}},
  \bibinfo{author}{\bibfnamefont{M.}~\bibnamefont{Rohlfing}},
  \bibinfo{author}{\bibfnamefont{P.}~\bibnamefont{Kr\"uger}}, \bibnamefont{and}
  \bibinfo{author}{\bibfnamefont{J.}~\bibnamefont{Pollmann}},
  \bibinfo{journal}{Phys. Rev. B} \textbf{\bibinfo{volume}{67}},
  \bibinfo{pages}{115111} (\bibinfo{year}{2003}),
  \urlprefix\url{https://link.aps.org/doi/10.1103/PhysRevB.67.115111}.

\bibitem[{\citenamefont{Comte and Mahler}(1986)}]{Comte_PhysRevB.34.7164}
\bibinfo{author}{\bibfnamefont{C.}~\bibnamefont{Comte}} \bibnamefont{and}
  \bibinfo{author}{\bibfnamefont{G.}~\bibnamefont{Mahler}},
  \bibinfo{journal}{Phys. Rev. B} \textbf{\bibinfo{volume}{34}},
  \bibinfo{pages}{7164} (\bibinfo{year}{1986}),
  \urlprefix\url{https://link.aps.org/doi/10.1103/PhysRevB.34.7164}.

\bibitem[{\citenamefont{Comte and Mahler}(1988)}]{Comte_PhysRevB.38.10517}
\bibinfo{author}{\bibfnamefont{C.}~\bibnamefont{Comte}} \bibnamefont{and}
  \bibinfo{author}{\bibfnamefont{G.}~\bibnamefont{Mahler}},
  \bibinfo{journal}{Phys. Rev. B} \textbf{\bibinfo{volume}{38}},
  \bibinfo{pages}{10517} (\bibinfo{year}{1988}),
  \urlprefix\url{https://link.aps.org/doi/10.1103/PhysRevB.38.10517}.

\bibitem[{\citenamefont{Latini et~al.}(0)\citenamefont{Latini, Ronca,
  De~Giovannini, H\"ubener, and Rubio}}]{Latini_QED-BSE}
\bibinfo{author}{\bibfnamefont{S.}~\bibnamefont{Latini}},
  \bibinfo{author}{\bibfnamefont{E.}~\bibnamefont{Ronca}},
  \bibinfo{author}{\bibfnamefont{U.}~\bibnamefont{De~Giovannini}},
  \bibinfo{author}{\bibfnamefont{H.}~\bibnamefont{H\"ubener}},
  \bibnamefont{and} \bibinfo{author}{\bibfnamefont{A.}~\bibnamefont{Rubio}},
  \bibinfo{journal}{Nano Letters} \textbf{\bibinfo{volume}{0}},
  \bibinfo{pages}{null} (\bibinfo{year}{0}),
  \urlprefix\url{https://doi.org/10.1021/acs.nanolett.9b00183}.

\bibitem[{\citenamefont{Perfetto and Stefanucci}(2018)}]{PS-cheers}
\bibinfo{author}{\bibfnamefont{E.}~\bibnamefont{Perfetto}} \bibnamefont{and}
  \bibinfo{author}{\bibfnamefont{G.}~\bibnamefont{Stefanucci}},
  \bibinfo{journal}{Journal of Physics: Condensed Matter}
  \textbf{\bibinfo{volume}{30}}, \bibinfo{pages}{465901}
  (\bibinfo{year}{2018}),
  \urlprefix\url{http://stacks.iop.org/0953-8984/30/i=46/a=465901}.

\bibitem[{\citenamefont{Perfetto et~al.}(2016)\citenamefont{Perfetto, Sangalli,
  Marini, and Stefanucci}}]{PSMS.2016}
\bibinfo{author}{\bibfnamefont{E.}~\bibnamefont{Perfetto}},
  \bibinfo{author}{\bibfnamefont{D.}~\bibnamefont{Sangalli}},
  \bibinfo{author}{\bibfnamefont{A.}~\bibnamefont{Marini}}, \bibnamefont{and}
  \bibinfo{author}{\bibfnamefont{G.}~\bibnamefont{Stefanucci}},
  \bibinfo{journal}{Phys. Rev. B} \textbf{\bibinfo{volume}{94}},
  \bibinfo{pages}{245303} (\bibinfo{year}{2016}),
  \urlprefix\url{https://link.aps.org/doi/10.1103/PhysRevB.94.245303}.

\bibitem[{\citenamefont{Freericks et~al.}(2009)\citenamefont{Freericks,
  Krishnamurthy, and Pruschke}}]{Freericks_PhysRevLett.102.136401}
\bibinfo{author}{\bibfnamefont{J.~K.} \bibnamefont{Freericks}},
  \bibinfo{author}{\bibfnamefont{H.~R.} \bibnamefont{Krishnamurthy}},
  \bibnamefont{and} \bibinfo{author}{\bibfnamefont{T.}~\bibnamefont{Pruschke}},
  \bibinfo{journal}{Phys. Rev. Lett.} \textbf{\bibinfo{volume}{102}},
  \bibinfo{pages}{136401} (\bibinfo{year}{2009}),
  \urlprefix\url{https://link.aps.org/doi/10.1103/PhysRevLett.102.136401}.

\bibitem[{\citenamefont{My\"oh\"anen et~al.}(2010)\citenamefont{My\"oh\"anen,
  Stan, Stefanucci, and van Leeuwen}}]{Myohanen_2010}
\bibinfo{author}{\bibfnamefont{P.}~\bibnamefont{My\"oh\"anen}},
  \bibinfo{author}{\bibfnamefont{A.}~\bibnamefont{Stan}},
  \bibinfo{author}{\bibfnamefont{G.}~\bibnamefont{Stefanucci}},
  \bibnamefont{and} \bibinfo{author}{\bibfnamefont{R.}~\bibnamefont{van
  Leeuwen}}, \bibinfo{journal}{Journal of Physics: Conference Series}
  \textbf{\bibinfo{volume}{220}}, \bibinfo{pages}{012017}
  (\bibinfo{year}{2010}),
  \urlprefix\url{https://doi.org/10.1088%2F1742-6596%2F220%2F1%2F012017}.

\bibitem[{\citenamefont{Lipavsk\'y et~al.}(1986)\citenamefont{Lipavsk\'y,
  \ifmmode \check{S}\else \v{S}\fi{}pi\ifmmode~\check{c}\else \v{c}\fi{}ka, and
  Velick\'y}}]{PhysRevB.34.6933}
\bibinfo{author}{\bibfnamefont{P.}~\bibnamefont{Lipavsk\'y}},
  \bibinfo{author}{\bibfnamefont{V.}~\bibnamefont{\ifmmode \check{S}\else
  \v{S}\fi{}pi\ifmmode~\check{c}\else \v{c}\fi{}ka}}, \bibnamefont{and}
  \bibinfo{author}{\bibfnamefont{B.}~\bibnamefont{Velick\'y}},
  \bibinfo{journal}{Phys. Rev. B} \textbf{\bibinfo{volume}{34}},
  \bibinfo{pages}{6933} (\bibinfo{year}{1986}),
  \urlprefix\url{https://link.aps.org/doi/10.1103/PhysRevB.34.6933}.

\end{thebibliography}

\end{document}